\documentclass[11pt,onecolumn,showpacs,floatfix,superscriptaddress,nofootinbib,amssymb,amsmath,aps,prc,eqsecnum]{revtex4-2}
\usepackage[dvips]{graphicx}
\usepackage{epstopdf} 
\usepackage{latexsym,amssymb,amsmath,epsfig,bm,times,psfrag,subfigure}
\usepackage{color}
\usepackage{dcolumn}                 % Align table columns on decimal point
\usepackage[bookmarksnumbered,bookmarksopen,colorlinks,citecolor=blue,linkcolor=blue]{hyperref}
\newcommand{\fig}[1]{Fig.~\ref{#1}}
\newcommand{\eq}[1]{Eq.~(\ref{#1})}

\renewcommand{\part}{{\rm part}}
\newcommand{\be}{\begin{equation}}
\newcommand{\ee}{\end{equation}}
\newcommand{\bear}{\begin{eqnarray}}
\newcommand{\eear}{\end{eqnarray}}
\newcommand{\ba}{\begin{array}}
	\newcommand{\ea}{\end{array}}

\begin{document}
\title{Electromagnetic Field Produced in High Energy Small Collision System within Charge Density Models of Nucleon}

\author{Zong-Wei Zhang}
\affiliation{School of Physics, Huazhong University of Science and Technology, Wuhan 430074, China}

\author{Xian-Zhuo Cen}
\affiliation{School of Physics, Huazhong University of Science and Technology, Wuhan 430074, China}
\affiliation{Dalingshan Secondary School, DongGuan 523000, China}

%\author{Xu-Guang Huang}
%\affiliation{Physics Deparment and Center for Particle Physics and Field Theory, Fudan Univeristy, Shanghai 200433, China}

\author{Wei-Tian Deng}
\email{dengwt@hust.edu.cn}
\affiliation{School of Physics, Huazhong University of Science and Technology, Wuhan 430074, China}

%\date{\today}

\begin{abstract}
Recent experiments show that $\Delta\gamma$, an observable designed for detecting the chiral magnetic effect (CME), in small collision system $p+A$ is similar with that in heavy ion collision $A+A$. This brings a challenge to the existence of CME because it is believed that there is no azimuthal correlation between the orientation of the magnetic field ($\Phi_B$) and the participant plane ($\Phi_2$) in small collision system. In this work, we introduce three charge density models to describe the inner charge distributions of proton and neutron, and calculate the electric and magnetic fields produced in small $p+A$ collisions at both RHIC and LHC energies. Our results show that the contribution of the single projectile proton to the magnetic field is the main source after average over all participants. The azimuthal correlation between $\Phi_B$ and $\Phi_2$ is small but not vanished. And due to the huge fluctuation of fields strength, the magnetic-field contribution to $\Delta\gamma$ could be large. 
\end{abstract}
\maketitle

\section {introduction}\label{section:discu}

In high energy heavy ion collisions, quark-gluon-plasma (QGP) is produced due to the extreme environment of high temperature and high pressure. Since the colliding large ions have positive electric charge, a strong transient electric and magnetic fields are also produced \cite{Rafelski:1975rf,Skokov:2009qp,Bzdak:2011yy,Voronyuk:2011jd,Deng:2012pc,Deng:2014uja,Inghirami:2016iru,Yan:2021zjc} in off-central collisions. This extremely strong fields provide a unique environment to study properties of quantum chromodynamics (QCD). People have realized that there maybe is a parity symmetry($\mathcal{P}$) or charge conjugate and parity symmetry($\mathcal{CP}$) violation effect in QCD~\cite{Kharzeev:1998kz,Kharzeev:2004ey,Kharzeev:2007jp,Fukushima:2008xe,Kharzeev:2015znc,Liu:2020ymh}. This effect could be observed as chiral magnetic effect (CME) once coupled to a strong magnetic field~\cite{Kharzeev:2007jp,Fukushima:2008xe}. The high energy heavy-ion collision could give an excellent environment for studying CME, because not only the QGP with high temperature could be generated but also an extremely strong magnetic field is produced.  Search of CME is one of the most important tasks of high energy heavy ion collision physics. The measurement of the charge separation phenomenon induced by CME can provide a way to study the quantum anomaly of QCD vacuum topology. 

However, the charge separation phenomenon cannot be directly observed, so a three-point correlator
\begin{equation}
  \gamma=<\cos(\alpha+\beta-2\Phi_2)>
\end{equation}
 was proposed~\cite{Voloshin:2004vk}, where $\alpha$ and $\beta$ is azimuthal angles of charged particle, the $\Phi_2$ is the angle of the reaction plane for a given case. Significant charge distribution anisotropy $\Delta\gamma$
  \begin{equation}
 	\Delta\gamma\equiv\gamma_{OS}-\gamma_{SS}
 \end{equation}
 has indeed been measured in heavy-ion collision experiments~\cite{Abelev:2009ac,Abelev:2009ad,Adamczyk:2014mzf,Sirunyan:2017quh,Acharya:2017fau} which show features consistent with the CME expectation. Here, $\gamma_{OS}$ represents for opposite charge particle pair, $\gamma_{SS}$ represents for same charge particle pair.
However, this observable may include the effect induced by elliptic-flow ($v_2$) induced backgrounds~\cite{Xu:2017zcn,Adamczyk:2013kcb,Wang:2016iov,Ajitanand:2010rc,Bzdak:2011np,Zhao:2017nfq}.  The CME and $v_2$-related background are driven by different physical mechanisms: the CME is very closely related to magnetic field, but $v_2$-related background is connected to the participant plane: $\Phi_2$.

The charge anisotropy is related not only the strength of fields, but also the azimuthal correlation between $\Phi_B$ and $\Phi_2$ ~\cite{Bloczynski:2012en,Bloczynski:2013mca}
 \begin{equation}
	\Delta\gamma\propto B^2\cdot\cos2(\Phi_B-\Phi_2)
\end{equation}
where B is magnetic field, $\Phi_B$ is the azimuthal angle of the magnetic field, $\Phi_2$ is the second-order corresponding initial event-plane angles.

People used to think of there should be no charge separation caused by the CME effect in small systems collision due to the absence of azimuthal correlation between $\Phi_B$ and $\Phi_2$ \cite{Zhao_2018, Belmont:2016oqp}. In recent years experiment data show that the results of $\Delta\gamma$ in small system collision is very similar with heavy nuclear collision, such as Au+Au and Pb+Pb \cite{Sirunyan:2017quh,Khachatryan:2016got}, which implies that the main contribution of the charge anisotropy ($\Delta\gamma$) in $A+A$ collisions comes from the elliptical flow background ($v_2$) but not CME.

In order to clarify the contribution of CME, it is very necessary to study the nature of magnetic fields in small collision systems. The aim of this work is to give a more clear study on structure of event-by-event generated electromagnetic fields in small systems considering inner charge distribution of a nucleon. we focus on both the magnitude of the magnetic field $\bf{B}$ and its azimuthal correlation $\left\langle \cos2(\Phi_B - \Phi_{2}) \right\rangle$. Furthermore, we will show that  $B^2\cdot\cos2(\Phi_B-\Phi_2)$ have a considerable value in small system.

We use the Lienard-Wiechert potentials to calculate the electric and magnetic fields:\\
\begin{eqnarray}
	\label{eq:Lienard-Wiechert}
	\nonumber
	\emph{e}\bf{E}{(t,\bf{r})}&=&\frac{\emph{e}^2}{4\pi} \sum_{n}\emph{Z}_\emph{n}(\bf{R})\frac{\bf{R}_\emph{n}-\emph{R}_\emph{n}\bf{v}_\emph{n}}{(\emph{R}_\emph{n}-\bf{R}_\emph{n}{\cdot}\bf{v}_\emph{n})^3}(1-\upsilon_\emph{n}^2)\\
	\emph{e}\bf{B}{(t,\bf{r})}&=&\frac{\emph{e}^2}{4\pi} \sum_{n}\emph{Z}_\emph{n}(\bf{R})\frac{\bf{v}_\emph{n}\times\bf{R}_\emph{n}}{(\emph{R}_\emph{n}-\bf{R}_\emph{n}{\cdot}\bf{v}_\emph{n})^3}(1-\upsilon_\emph{n}^2)
\end{eqnarray}
where $\bf{R}_\emph{n}=\bf{r} - \bf{r}_\emph{n}$ is the relative position of the  field point $\bf{r}$ to the source point $\bf{r}_\emph{n}$, and $\bf{r}_\emph{n}$ is the location of the $\emph{n}$th particle with velocity ${v}_\emph{n}$ at the retarded time $t_n =t-| \bf{r}- \bf{r}_n|$ \cite{Deng:2012pc}. The summations run over all charged particles in the system. Some theoretical uncertainties come from the modeling of the proton: treating the proton as a point charge or as a uniformly charged ball \cite{Deng:2016knn}.

In paper \cite{Zhao_2018}, the authors have calculated the electromagnetic fields produced in small system, treating the nucleon as a point-like charge particle with a distance cut-off to avoid possible large fluctuation. In paper \cite{Belmont:2016oqp}, the authors use a Gaussian function with $\sigma =0.4$ fm to simulate the inner charge distribution of proton. Their results show that the magnetic field direction and the eccentricity orientation are uncorrelated. However, these models could elimilate huge fluctuation of fields strength effectively, but they are not good description of realistic inner charge density of nucleon and loss the possible correlation between  $\Phi_B$ and $\Phi_2$  . We will show that in this paper, while considering realistic charge density inside a nucleon, the electricmagnetic fields produced by a  flying single nucleon could be more complicated, and lead to an azimuthal correlation between $\Phi_B$ and $\Phi_2$ in small collisions.
In this work, we use three different charge density models to discribe the proton and neutron, which are Point-Like model, Hard-Sphere model, and more physical Charge-Profile model.

This paper is organized as following: In section~\ref{section:charge-models} we introduce three different charge density models of proton and neutron. In section~\ref{section:field-single-nucleon}, we give the results of magnetic filed produced by a single nucleon with RHIC energy observed in a lab reference. In section~\ref{section:collision-geometry} and section~\ref{section:results}, we began to calculate the magnetic field produced in high energy small collision p+A.

\section{Charge distribution model of nucleon}\label{section:charge-models}
In the calculation of electromagnetic field produced in high energy ion collisions, people used to treat nucleon as a Point-Like particle with a charge $+e$ for proton and $0$ for neutron. This is the simplest simplification, and is a good approximation for heavy ion collisions. Because there are dozens of protons in these heavy ion collision system, the field produced by these nucleons are averaged among them, so the discrepancy from the charge profile of nucleon could be negligible.

However, this Point-Like model will bring a large numerical fluctuation to results in high energy ion collisions, especially for small collision system. In order to eliminate this numerical fluctuation, we introduce the second charge model for proton, Hard-Sphere model. In this model, the proton is treated as a sphere with homogeneous charge density.
\begin{equation}
\label{eq:Hard-Sphere}
\rho_\emph{p}(\emph{r})=
\begin{cases}
\ \rho_0 &\emph{r}\leq \emph{R}\\
\ 0 &\emph{r}>\emph{R}
\end{cases}
\end{equation}
with $\rho_0$ $\approx$ 0.354 fm$^{-3}$, $\emph{R}$ $\approx$ 0.88 fm.
While neutron is neutral, there should not be contribution to field from neutron within Point-Like model and Hard-Sphere model,  so we didn't include neutrons into our calculation framework within  these two models.

The Point-Like model and Hard-Sphere model are both simplifications for realistic inner charge distribution of proton. In order to discuss the field produced in small system more precisely, we need the physical 3-dimensional charge profile of nucleons, not only proton but also neutron, noted as Charge-Profile model.

\begin{figure}
	\centering
	\includegraphics[scale=0.5]{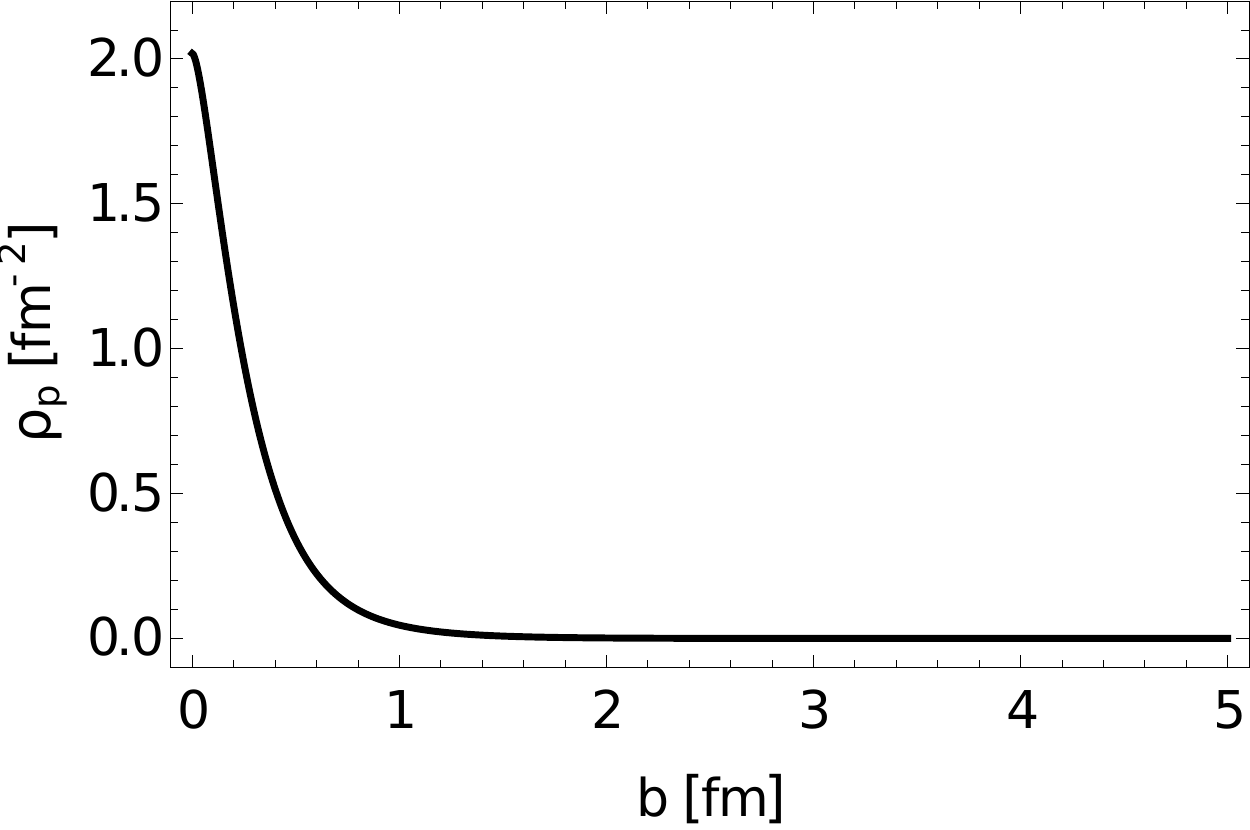}
	\includegraphics[scale=0.5]{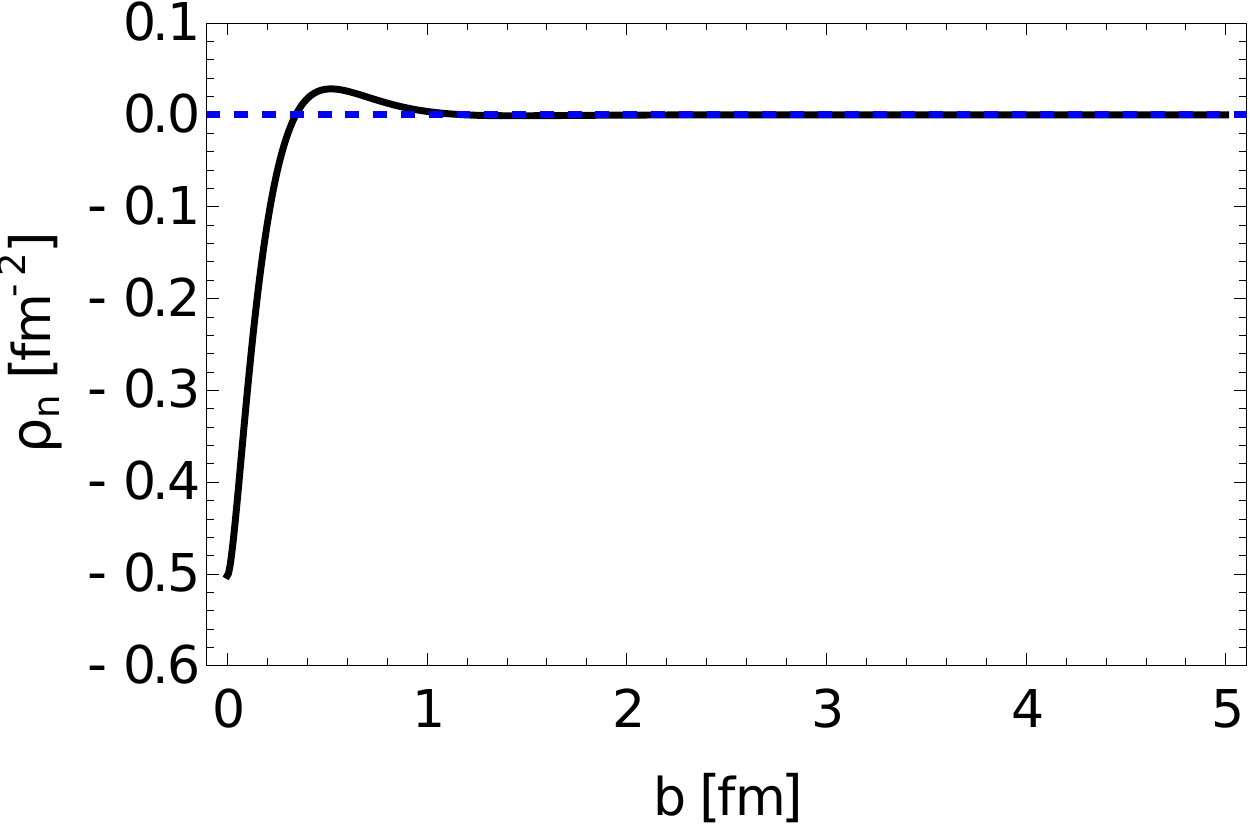}
	\caption{\label{fig:2Dcharge}The transverse charge density of proton (left) and neutron (right) in Charge-Profile model.}
\end{figure}

Since the transverse charge density of proton and neutron are available in \cite{Miller:2010nz}:
\begin{equation}
\label{eq:2D-rho}
\rho(\emph{b})=\int_{0}^{\infty} \frac{\emph{d}\emph{Q}}{2\pi}\emph{J}_0(\emph{Qb})\frac{\emph{G}_\emph{E}(\emph{Q}^2)+\tau\emph{G}_\emph{M}(\emph{Q}^2)}{1+\tau}
\end{equation}
with $\tau$ = $\frac{\emph{Q}^2}{4\emph{M}^2}$, $\emph{M}$ the mass of proton or neutron, $\emph{J}_0$ a cylindrical Bessel function. Employing the parameterization of electric form factor $\emph{G}_\emph{E}$ and magnetic form factor $\emph{G}_\emph{M}$\cite{Alberico:2008sz} into \eq{eq:2D-rho}, we can get the transverse charge density of proton and neutron shown in \fig{fig:2Dcharge}.

\begin{figure}
	\centering
	\includegraphics[scale=0.5]{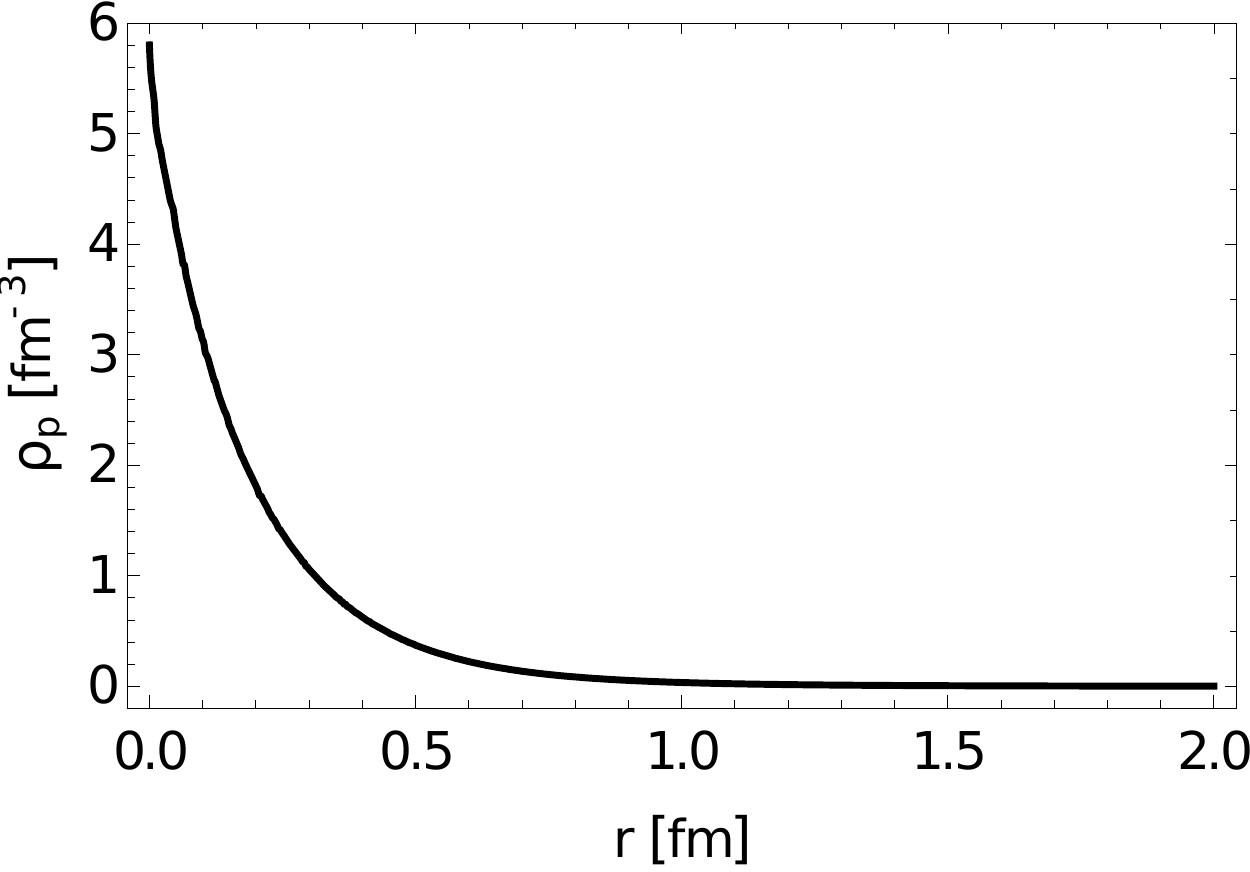}
	\includegraphics[scale=0.52]{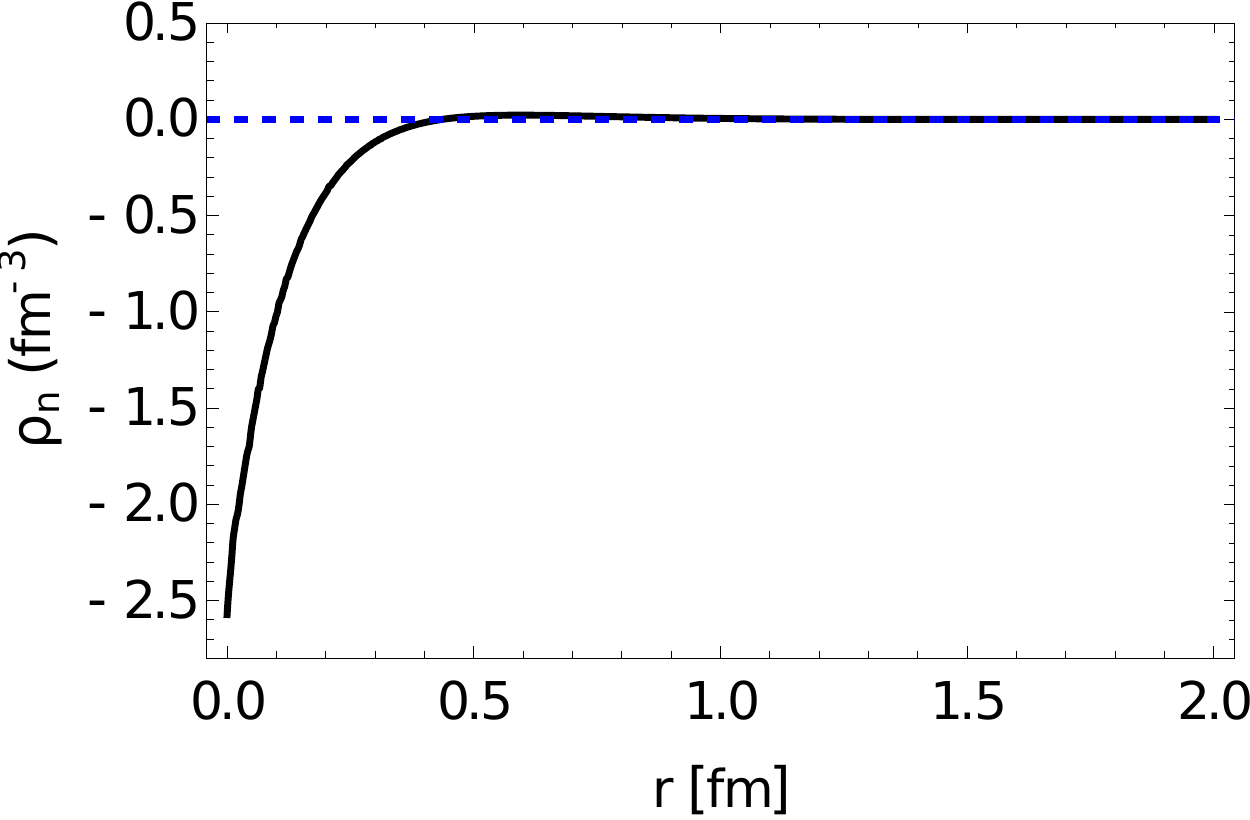}
	\caption{\label{fig:3Dcharge}The 3-dimensional charge profile of proton (left) and neutron (right) in Charge-Profile model.}
\end{figure}

Based on the transverse charge density, we can get the 3-dimensional charge profile of proton and neutron after inverted convolution with a spherical symmetry approximation, shown in  \fig{fig:3Dcharge}

\section{Magenic field produced by a single nucleon}\label{section:field-single-nucleon}

\begin{figure}
	\centering
	\subfigure{
		\label{fig:field-singleP-z0}
		\includegraphics[scale=0.6]{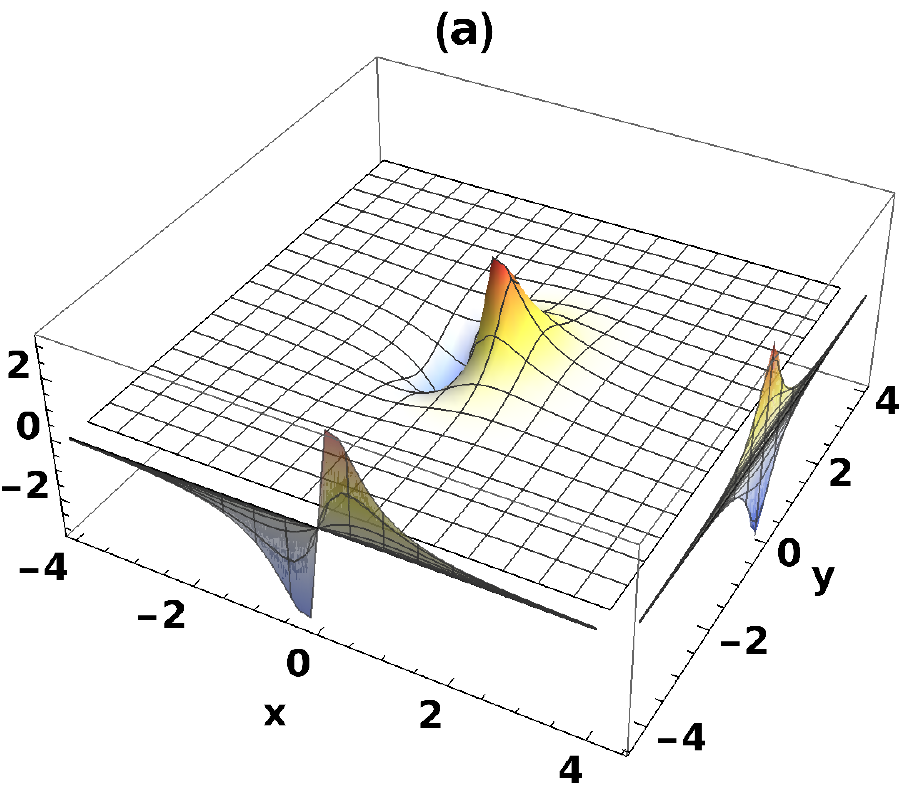}}
	\subfigure{
		\label{fig:field-singleP-z1}
		\includegraphics[scale=0.6]{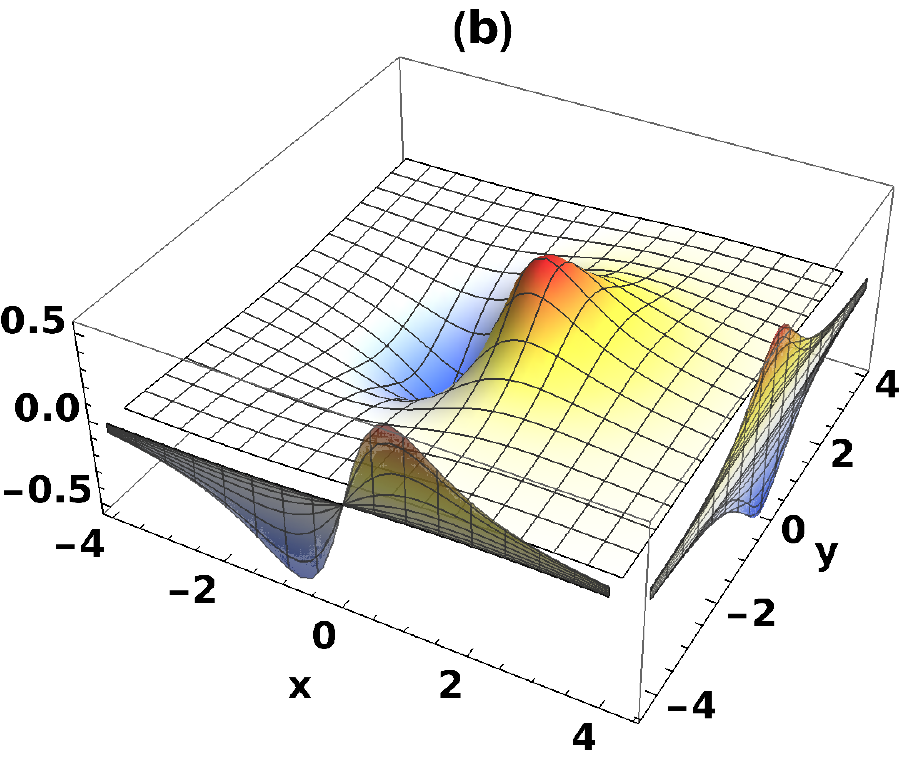}}
	\subfigure{
		\label{fig:field-singleP-z5}
		\includegraphics[scale=0.6]{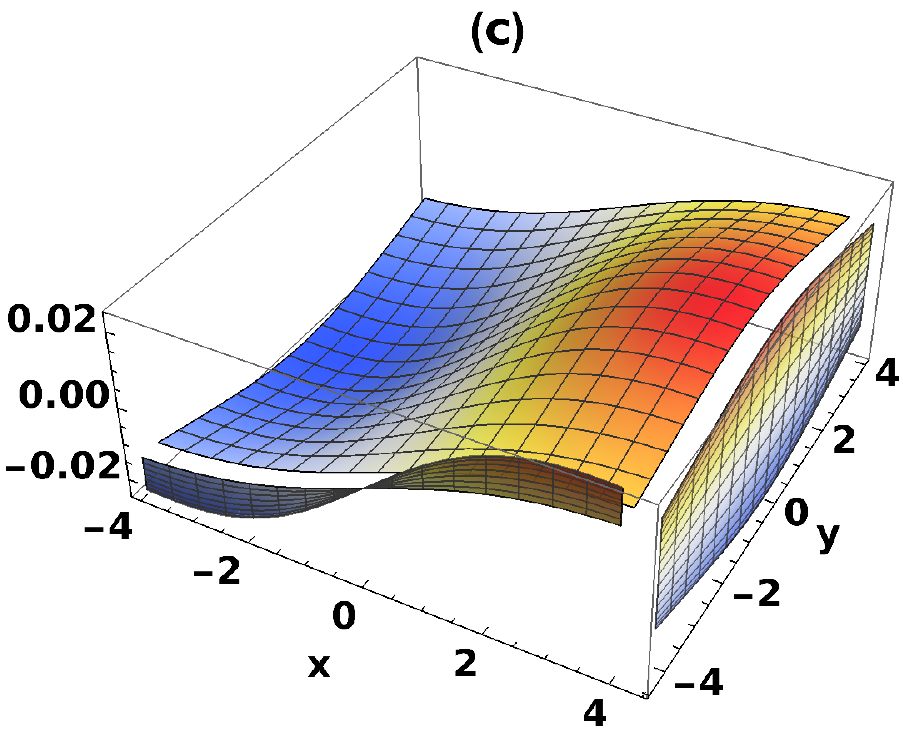}}
	\subfigure{
		\label{fig:field-singleP-z10}
		\includegraphics[scale=0.6]{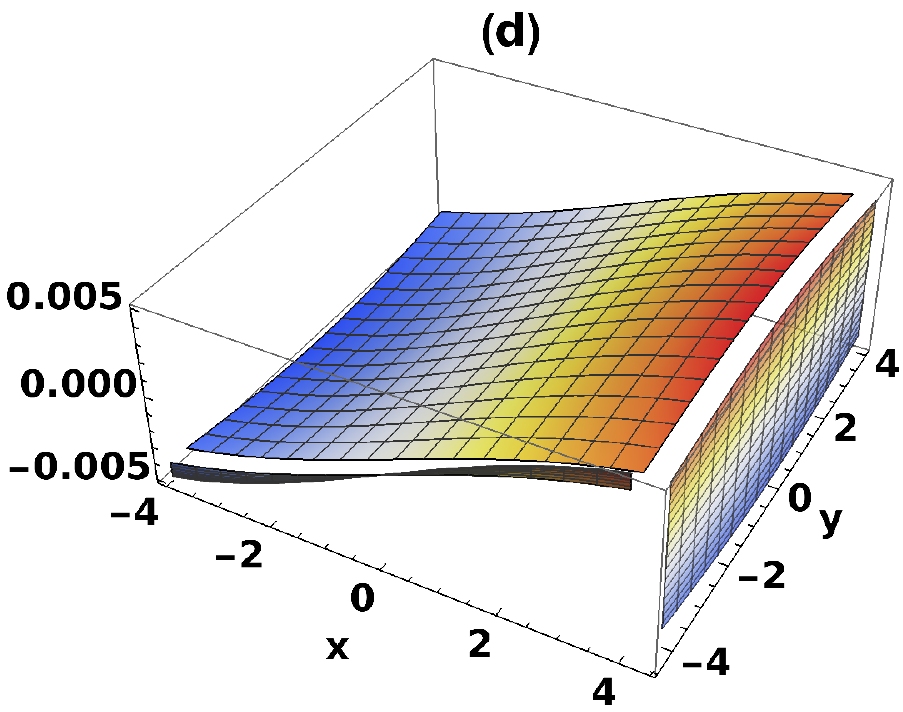}}
	\caption{\label{fig:field-singleP}The space distribution of magnetic field $B_y$ on $x-y$ plane with $z=0$ on \fig{fig:field-singleP-z0}, $z=1/\gamma$ (fm) on \fig{fig:field-singleP-z1}, $z=5/\gamma$ (fm) on \fig{fig:field-singleP-z5}, and $z=10/\gamma$ (fm) on \fig{fig:field-singleP-z10} produced by a single proton with 100 GeV travels through $+z$ direction, calculated within Charge-Profile model. }
\end{figure}

With inner charge distribution models, we can investigate the space distribution of field produced by a single nucleon  replacing the sum in  \eq{eq:Lienard-Wiechert} as integration.
Given a single proton with 100 GeV which is corresponding to RHIC energy, traveling along $+z$ direction. With the 3-dimensional Charge-Profile model, the results are given in \fig{fig:field-singleP}.  Since this single proton travels with an ultra-relativistic speed, the space distribution of field is Lorentz contracted with a factor $\gamma = E/m$. So we show our results on $x-y$ plane with four different z locations.

 In the first panel of \fig{fig:field-singleP}, the $x-y$ distribution of $B_y$ at $z=0$ is shown. We can see the magnetic field at the central location $r=(0,0,0)$, just the center of this single proton, is not divergence but 0 exactly. While we increase the value of $z$, the strength of $B_y$ is decreased. Shown on \fig{fig:field-singleP-z10}, we see the strength of $B_y$ at $z=10/\gamma$ fm is about  three order smaller than that on $z=0$.

\begin{figure}
	\centering
	\subfigure{
		\label{fig:field-singleP-PL-z0}
		\includegraphics[scale=0.6]{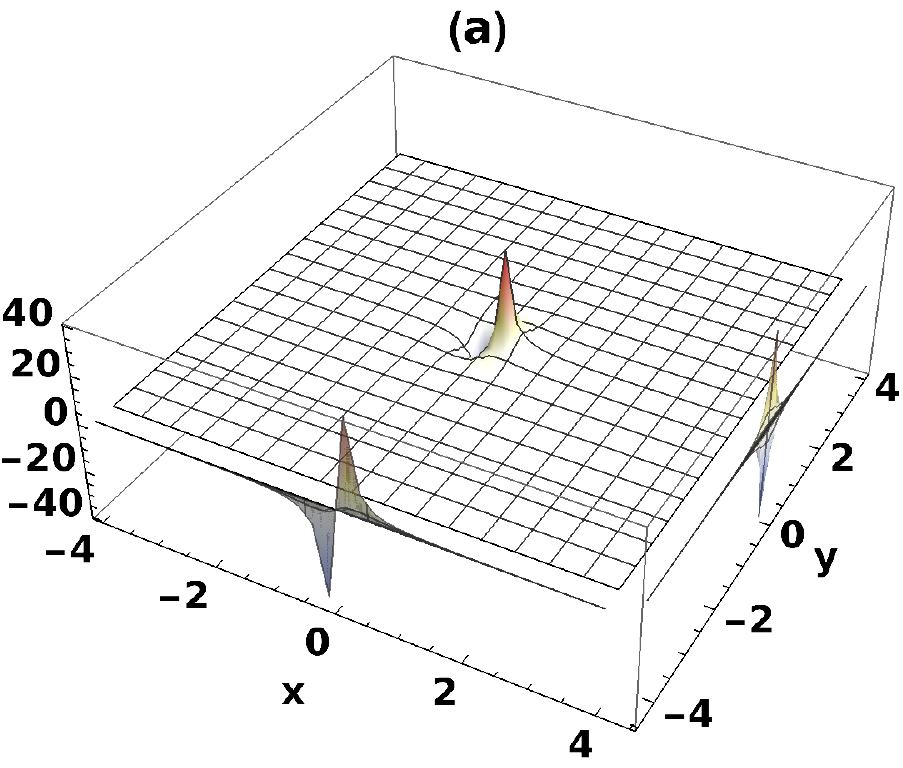}}
	\subfigure{
		\label{fig:field-singleP-PL-z1}
		\includegraphics[scale=0.6]{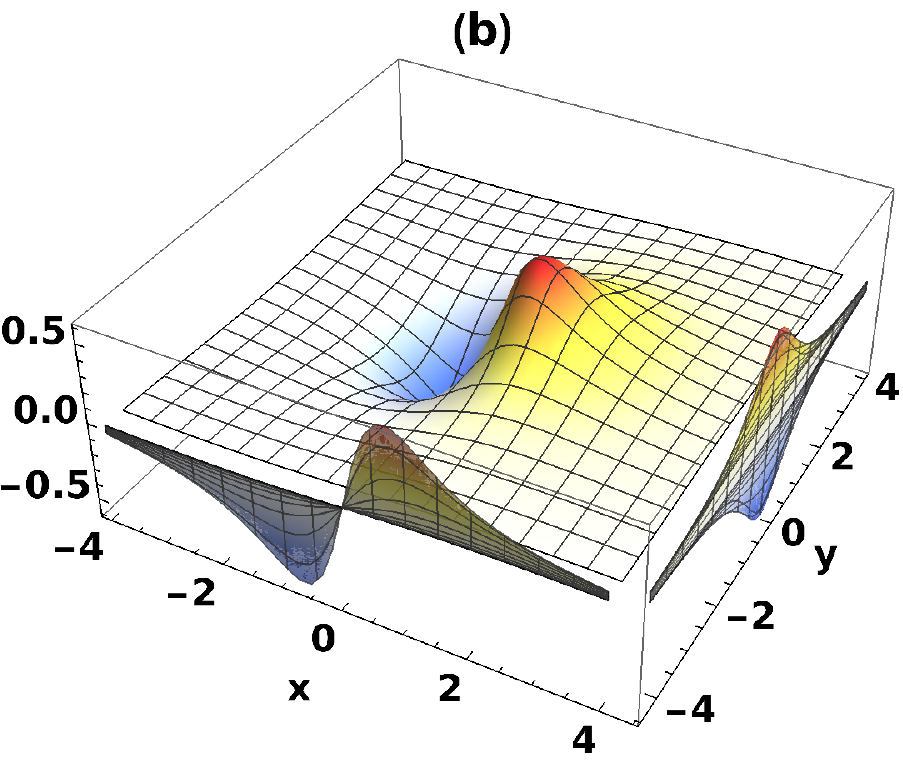}}
	\subfigure{
		\label{fig:field-singleP-PL-z5}
		\includegraphics[scale=0.6]{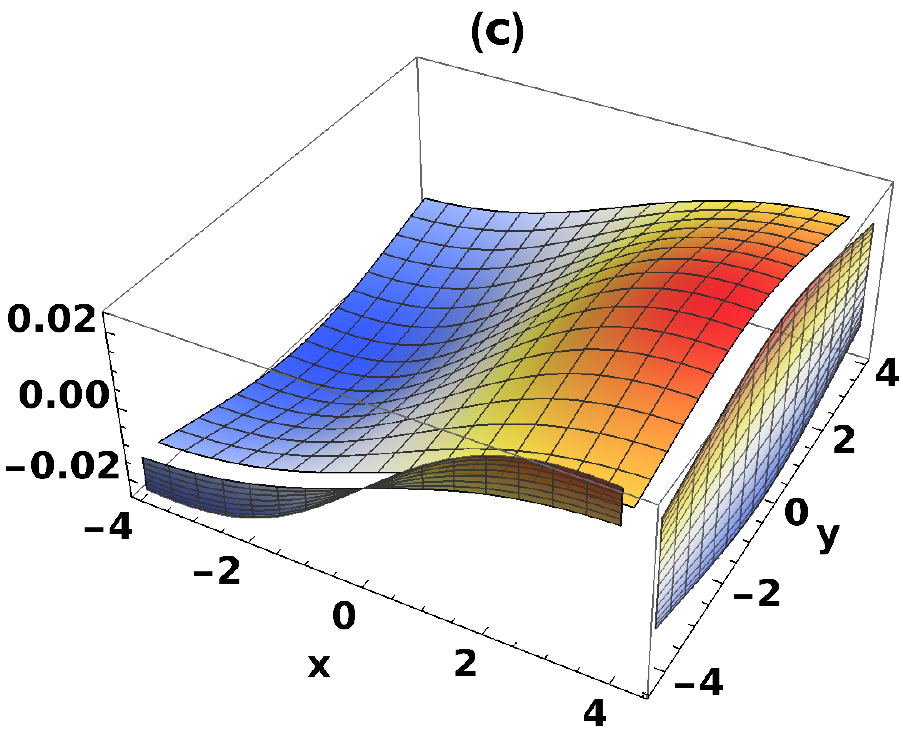}}
	\subfigure{
		\label{fig:field-singleP-PL-z10}
		\includegraphics[scale=0.6]{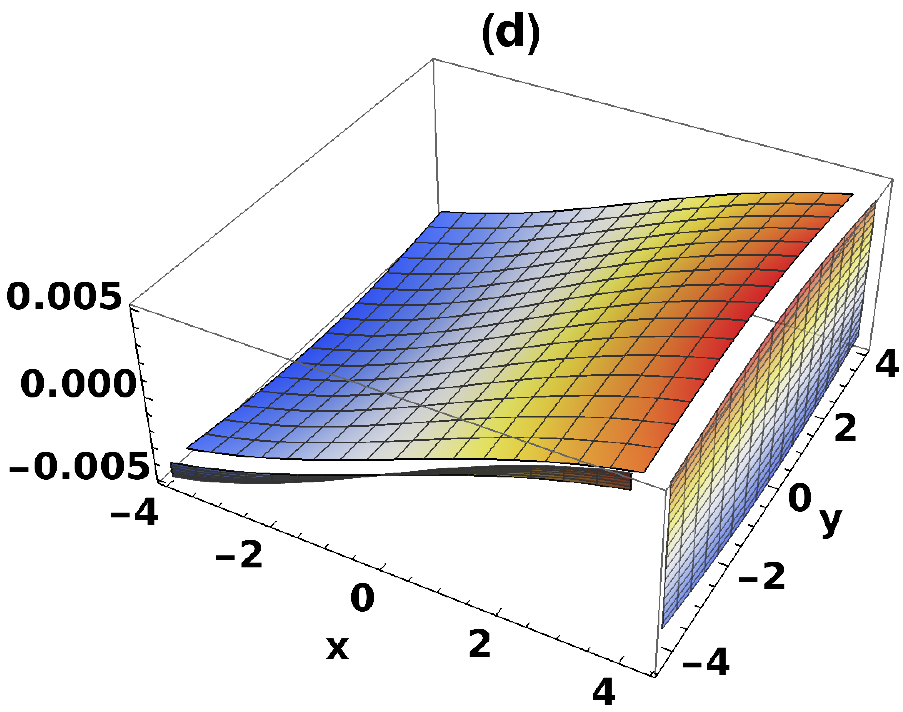}}
	\caption{\label{fig:field-singleP-PLmodel}The space distribution of magnetic field $B_y$ on $x-y$ plane with $z=0$ on \fig{fig:field-singleP-PL-z0}, $z=1/\gamma$ (fm) on \fig{fig:field-singleP-PL-z1}, $z=5/\gamma$ (fm) on \fig{fig:field-singleP-PL-z5}, and $z=10/\gamma$ (fm) on \fig{fig:field-singleP-PL-z10} produced by a single proton with 100 GeV travels through $+z$ direction, calculated within Point-Like model}
\end{figure}

Also, we checked the strength of  $B_y$ with Point-Like model. The corresponding results are shown on \fig{fig:field-singleP-PLmodel}. In this calculation, we have abandoned the result at $r=(0,0,0)$ in order to eliminate divergence. We can find that the strength of  $B_y$ in Point-Like model is much larger than what in Charge-Profile model at small value of $z$, while the strength of $B_y$ in the two models are similar at large $z$. From these calculation, we can see the results within Point-Like model could implement much larger fluctuation comparing with Charge-Profile model. For small collision system, because there are only a few nucleons have significant contribution to magnetic field, this fluctuation will bring us much trouble to handle on. And this fluctuation comes only from numerical calculation based on an unnecessary simplification.  That's why it's necessary to introduce the physical Charge-Profile model to calculate electromagnetic  field in this work.

\begin{figure}
	\centering
	\subfigure{
		\label{fig:field-singleN-z0}
		\includegraphics[scale=0.6]{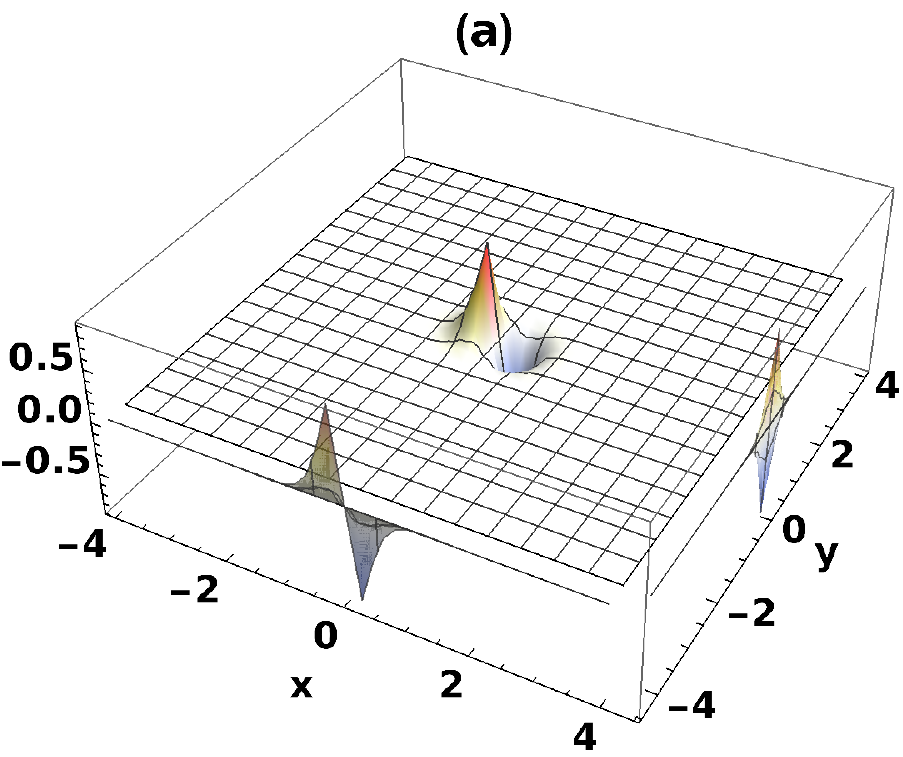}}
	\subfigure{
		\label{fig:field-singleN-z1}
		\includegraphics[scale=0.6]{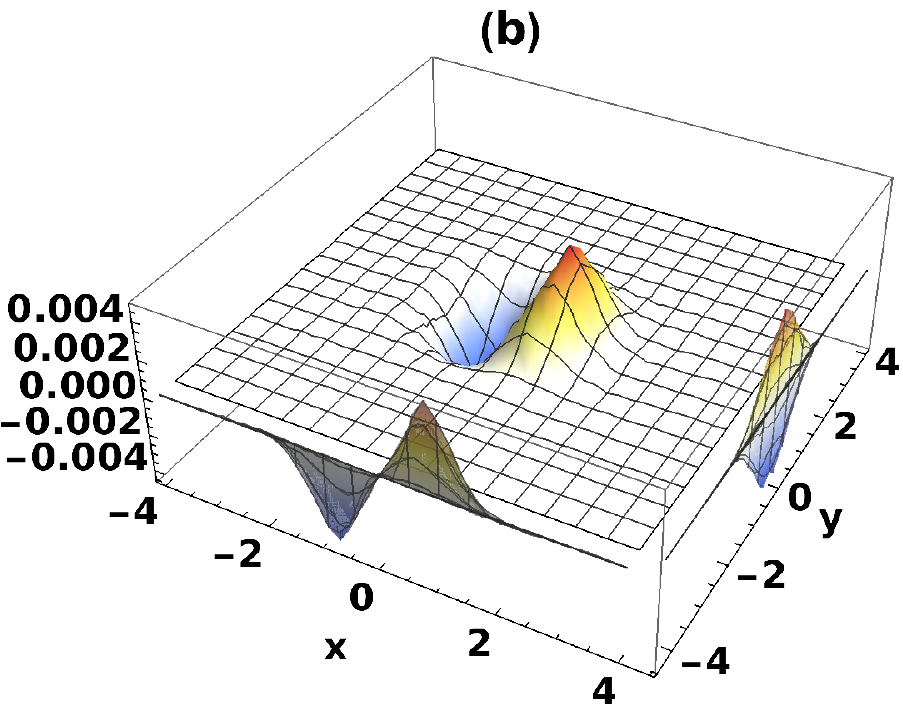}}
	\subfigure{
		\label{fig:field-singleN-z5}
		\includegraphics[scale=0.6]{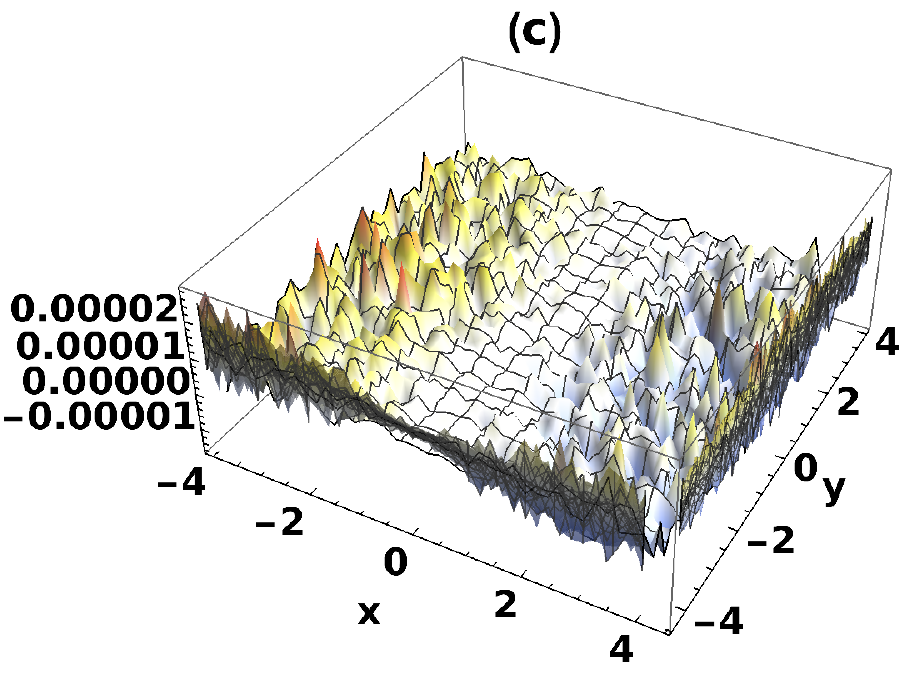}}
	\subfigure{
		\label{fig:field-singleN-z10}
		\includegraphics[scale=0.6]{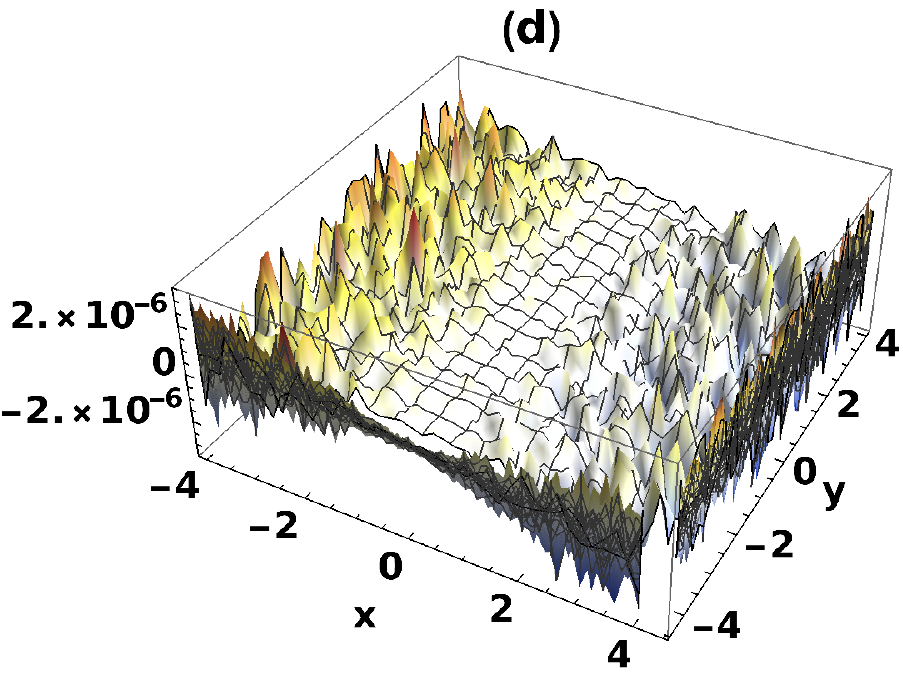}}
	\caption{\label{fig:field-singleN}The space distribution of magnetic field $B_y$ on $x-y$ plane with $z=0$ on \fig{fig:field-singleN-z0}, $z=1/\gamma$ (fm) on \fig{fig:field-singleN-z1}, $z=5/\gamma$ (fm) on \fig{fig:field-singleN-z5}, and $z=10/\gamma$ (fm) on \fig{fig:field-singleN-z10} produced by a single neutron with 100 GeV travels through $+z$ direction, calculated within Charge-Profile model.}
\end{figure}

In Point-Like model and Hard-Sphere model, neutron cannot contributed to electromagnetic field because of its neutral charge. But seeing from \fig{fig:3Dcharge}, neutron also has its charge profile. So neutron could also contribute to electromagnetic field in principle. Shown in \fig{fig:field-singleN}, the space distribution of $B_y$  is calculated within Charge-Profile model for neutron.  the strength of field is finite and significant smaller than  proton at small $z$ location, while it decrease rapidly to $0$ at large $z$ location. This is reasonable since neutron is charge neutral if it's measured from far away no matter what charge distribution in it.

\section{Geometry configuration of small collision system}\label{section:collision-geometry}
\begin{figure}[htbp]
	\centering
	\includegraphics[scale=0.35]{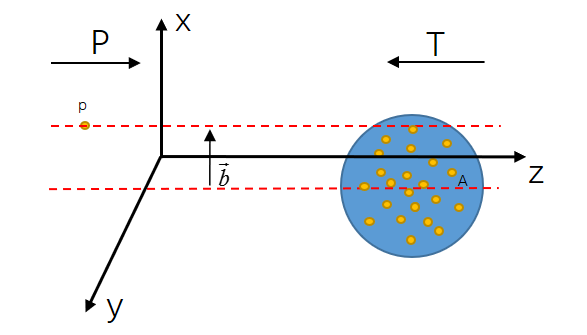}
	\includegraphics[scale=0.25]{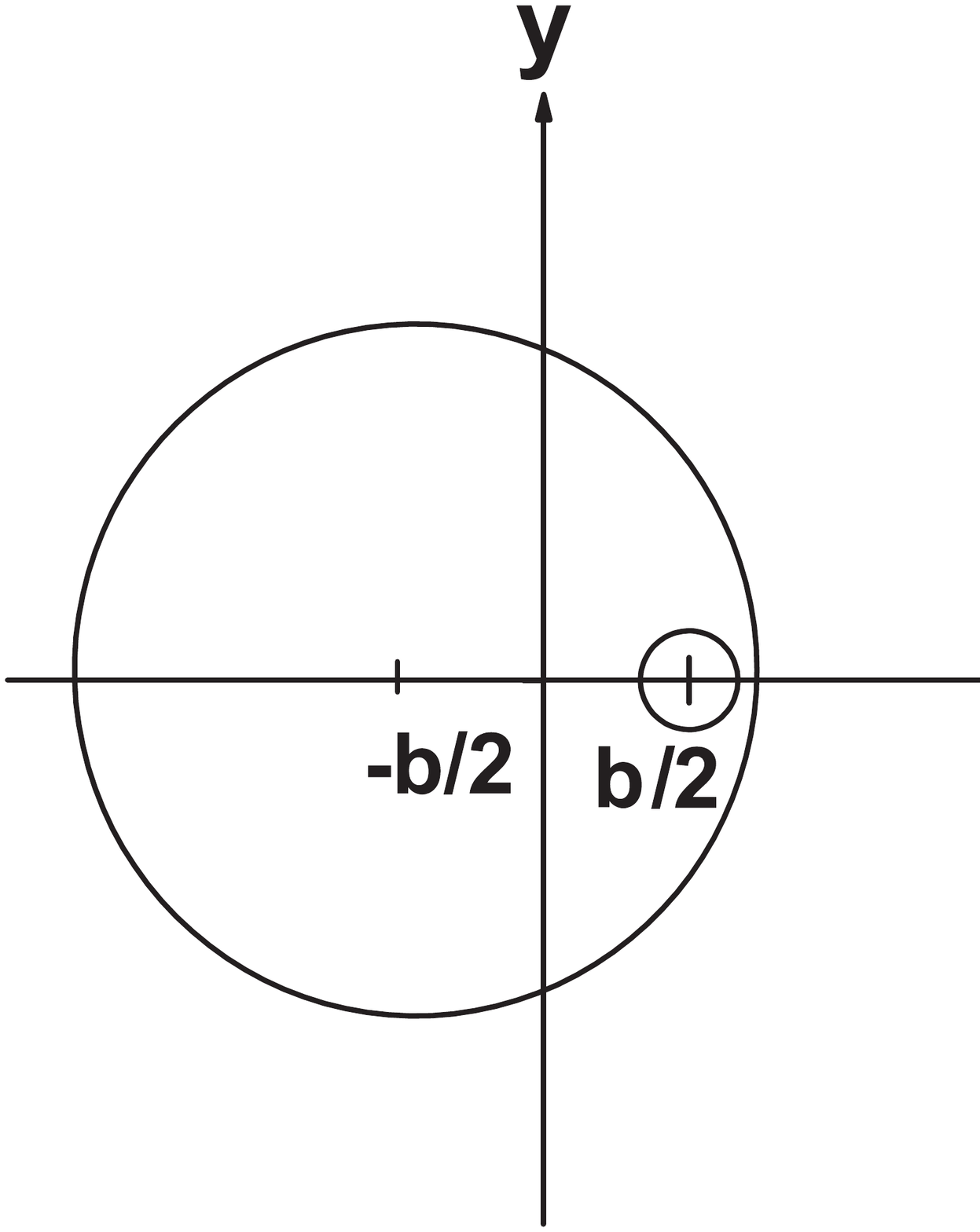}
	\caption{\label{fig:geometry-illustration}The geometry of small collision system $p+A$.}
\end{figure}

From this section, we concentrate on the electric and magnetic filed produced in high energy small collision system $p+A$. The collision geometry in our calculation is shown in \fig{fig:geometry-illustration}, where $b$ is the impact parameter which is the distance between the centers of two nuclear. In our framework, the single proton $p$ is treated as projectile travels along $+z$ direction, while the heavy ion $A$ is treated as target travels along $-z$ direction. we set $z$ coordination pass through the center of $b$, and put the projectile at $x=b/2$ and target at $x=-b/2$.

Inside the heavy ion, the position of each nucleon is sampled according to Woods-Saxon distribution
\begin{equation}
\rho{(r)}=\frac{\rho_0}{1+\exp[(r-R)/a]}
\end{equation}
where $\rho_0$=0.17 fm$^{-3}$, $a=$0.535 fm, $R$ is the radius of coming heavy ion.
Then the number of participants in each event  could be determined from Monte-Carlo Glauber model.

\section{Results and Discussions}\label{section:results}
To understand and analyze theoretical uncertainties, we will calculate the$\langle\emph{E}_\emph{x}\rangle$, $\langle|\emph{B}_\emph{x}|\rangle$, $\langle-\emph{B}_\emph{y}\rangle$ , and $\langle|\emph{B}_\emph{y}|\rangle$ using Point-Like model, Hard-Sphere model, and Charge-Profile model based on our available results of fields produced by single nucleon in section\ref{section:field-single-nucleon}. Here, the angle bracket is to denote event average. All results shown below are the strength of fields on location $\bf{r}_c$ which is the centre of mass of all participants.

\subsection{Impact parameter dependence of fields}

\begin{figure}[htbp]
\centering
	\subfigure{
		\label{fig:fields-pA-PL-Ex}
		\includegraphics[scale=0.6]{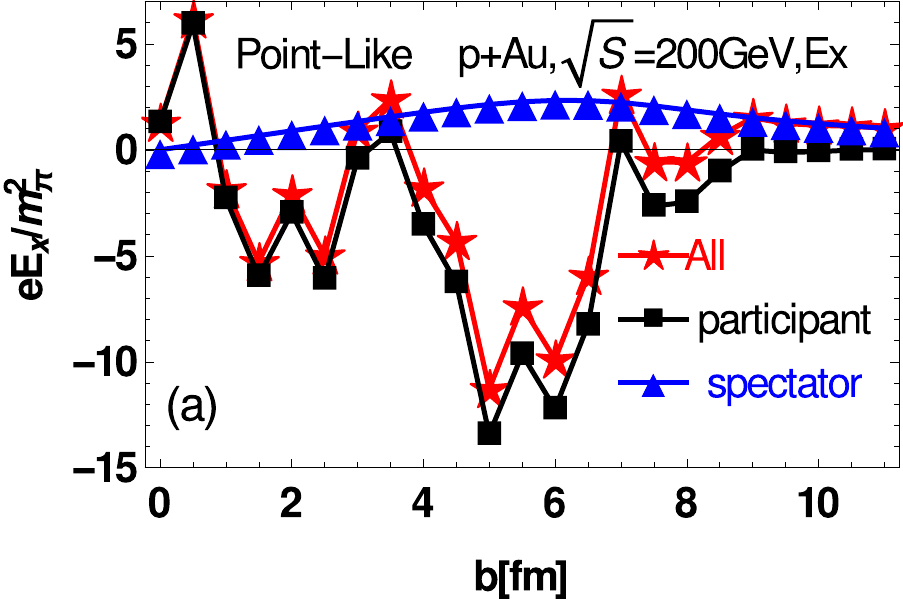}}
	\subfigure{
		\label{fig:fields-pA-PL-By}
		\includegraphics[scale=0.6]{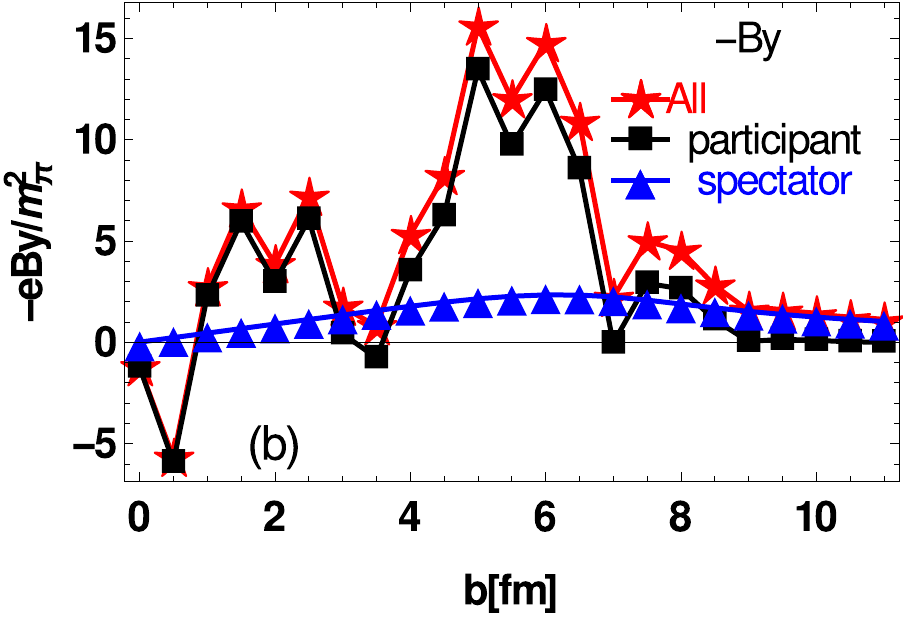}}
		\subfigure{
		\label{fig:fields-pA-PL-absBx}
		\includegraphics[scale=0.6]{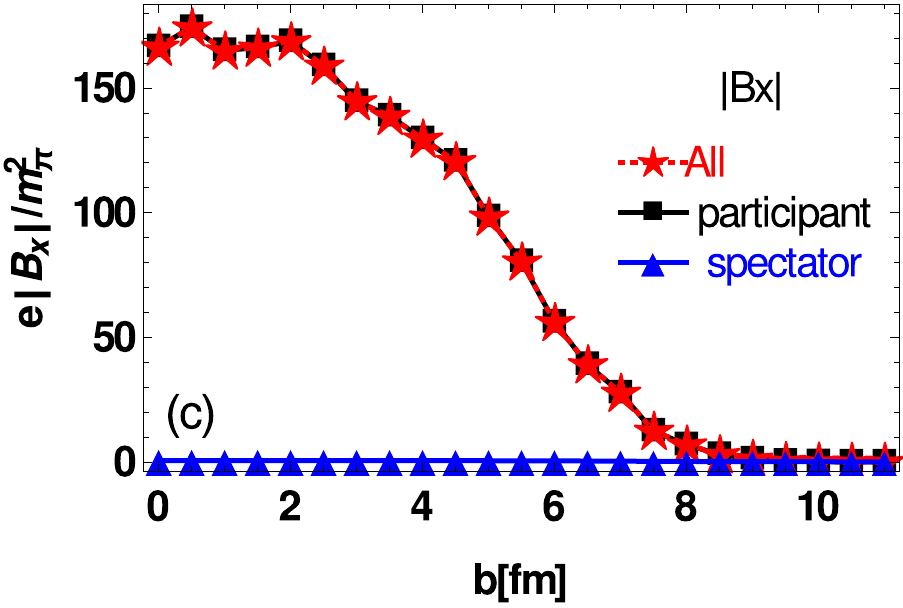}}
	\subfigure{
		\label{fig:fields-pA-PL-absBy}
		\includegraphics[scale=0.6]{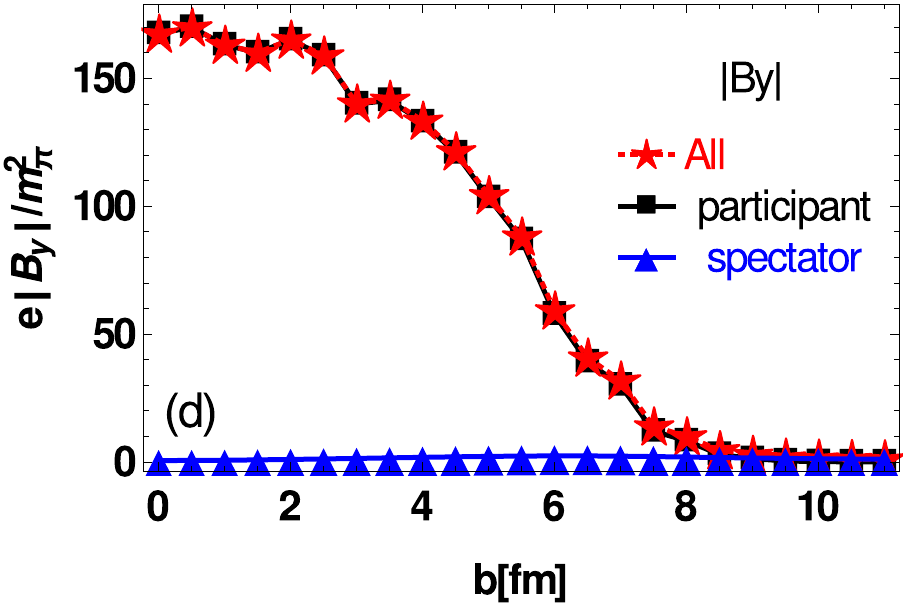}}
\caption{\label{fig:fields-pA-PL}The electromagnetic fields at $t$=0 and $\bf{r}=\bf{r}_c$ as function of the impact parameter b with Point-Like model.}
\end{figure}

Firstly, in \fig{fig:fields-pA-PL}, we show the  impact parameter dependence of the strength of fields produced within Point-Like model at RHIC energy.  After averaged over $10^6$ events, the value of $E_x$ and $B_y$ are shown (red curves in \fig{fig:fields-pA-PL-Ex} and \fig{fig:fields-pA-PL-By}) still with large fluctuation as we expected.  After checking contributions from participants nucleons and spectators nucleons, we find that this large fluctuations come from participants.

In small collision system $p+A$, there are only a few participants in each event due to the small size of single projectile proton. So these participants are all near to the observation point $\bf{r}_c$. The contribution from each participant  to fields could has a large strength as shown in \fig{fig:field-singleP-PLmodel}. But because the position of each nucleon is random inside nucleus, the orientation of  this field contributed from each participant is almost azimuthal random. Then the total field summed over these participants could remain a large strength but azimuthal random.  That's why there are so large fluctuations seen within Point-Like model.

In order to get the strength of fluctuation, we also calculate the impact parameter dependence of $|B_x|$ and $|B_y|$ shown on the \fig{fig:fields-pA-PL-absBx} and \fig{fig:fields-pA-PL-absBy}. Seen from these two panels, the averaged fluctuations of fields are very large, which is almost 80 times larger than fields strength in Au+Au collisions at RHIC energy\cite{Deng:2012pc}.

\begin{figure}[htbp]
\centering
	\subfigure{
		\label{fig:fields-pA-HS-Ex}
		\includegraphics[scale=0.6]{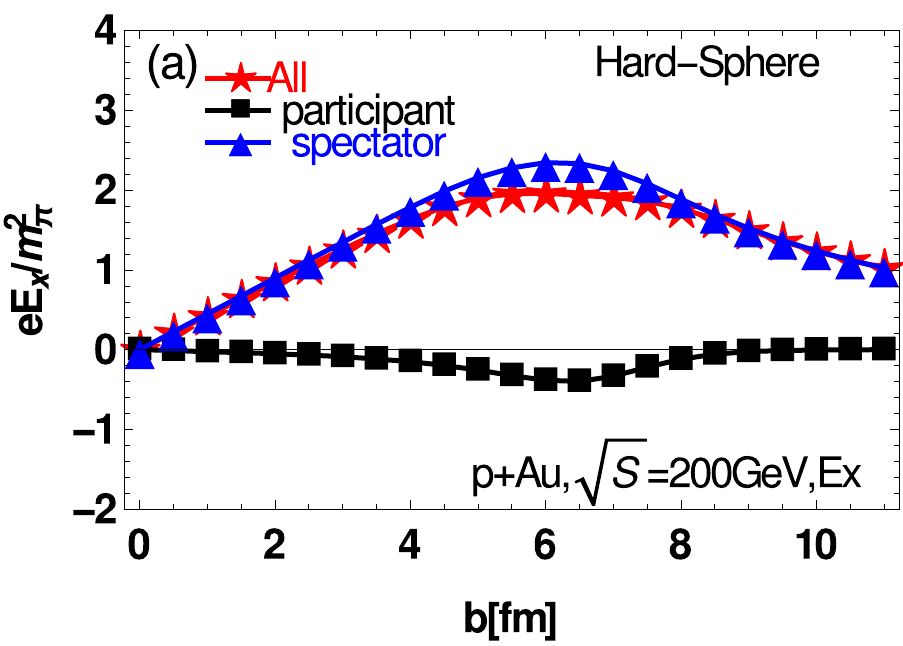}}
	\subfigure{
		\label{fig:fields-pA-HS-By}
		\includegraphics[scale=0.6]{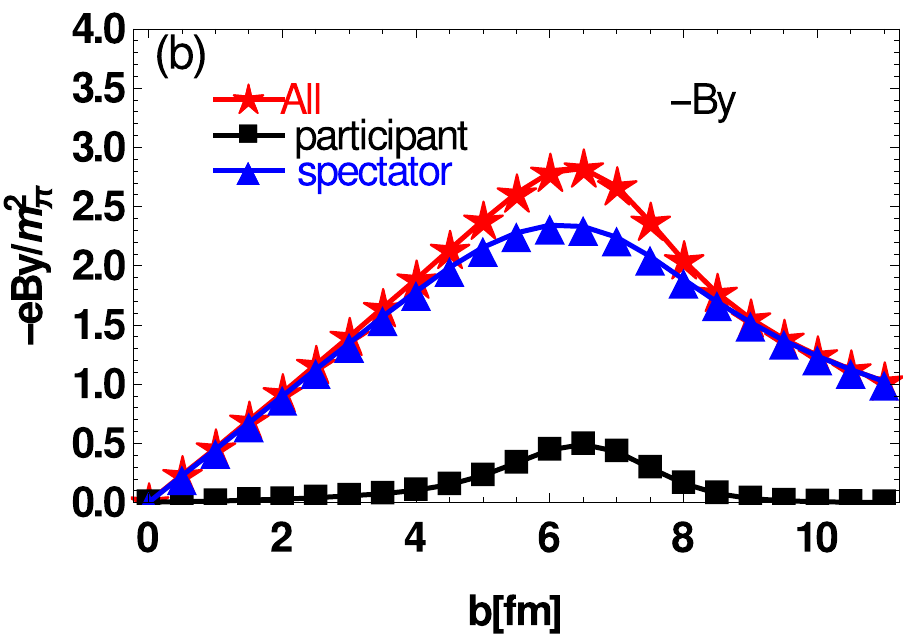}}
	\subfigure{
		\label{fig:fields-pA-HS-absBx}
		\includegraphics[scale=0.6]{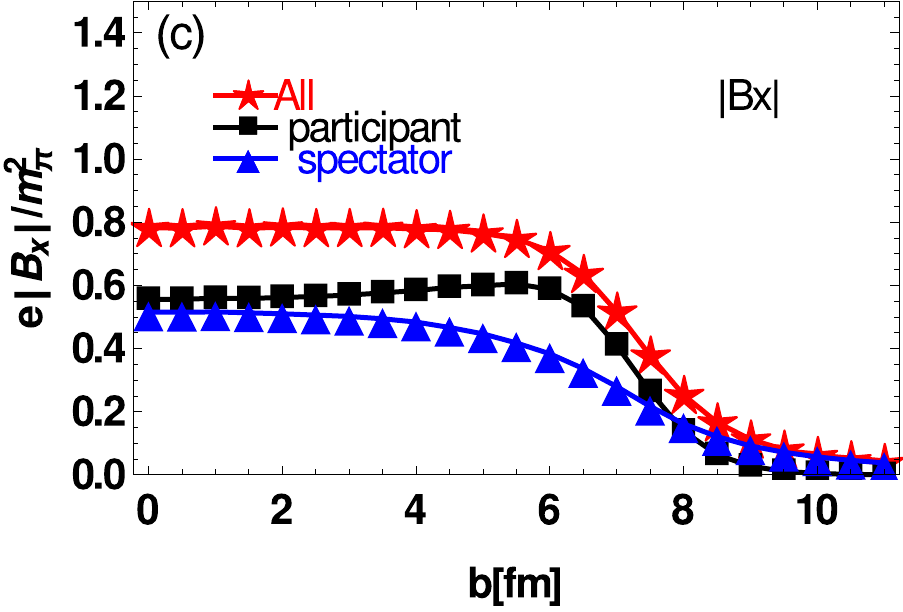}}
	\subfigure{
		\label{fig:fields-pA-HS-absBy}
		\includegraphics[scale=0.6]{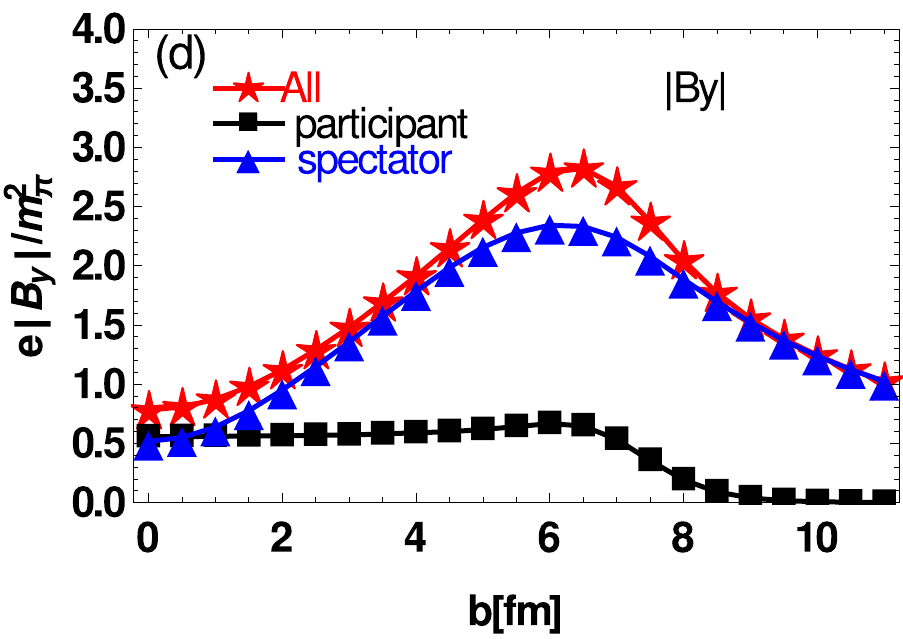}}
\caption{\label{fig:fields-pA-HS}The electromagnetic fields at $t$=0 and $\bf{r}=\bf{r}_c$ as function of the impact parameter b with Hard-Sphere model.}
\end{figure}

With Hard-Shpere model, the impact parameter dependence of fields are shown in \fig{fig:fields-pA-HS}. The curves in \fig{fig:fields-pA-HS} are much smoother than what in  \fig{fig:fields-pA-PL}. Seen from \fig{fig:fields-pA-HS-absBx} and \fig{fig:fields-pA-HS-absBy}, the fluctuations are also much smaller than what in Point-Like model.

\begin{figure}
		\centering
			\includegraphics[scale=0.6]{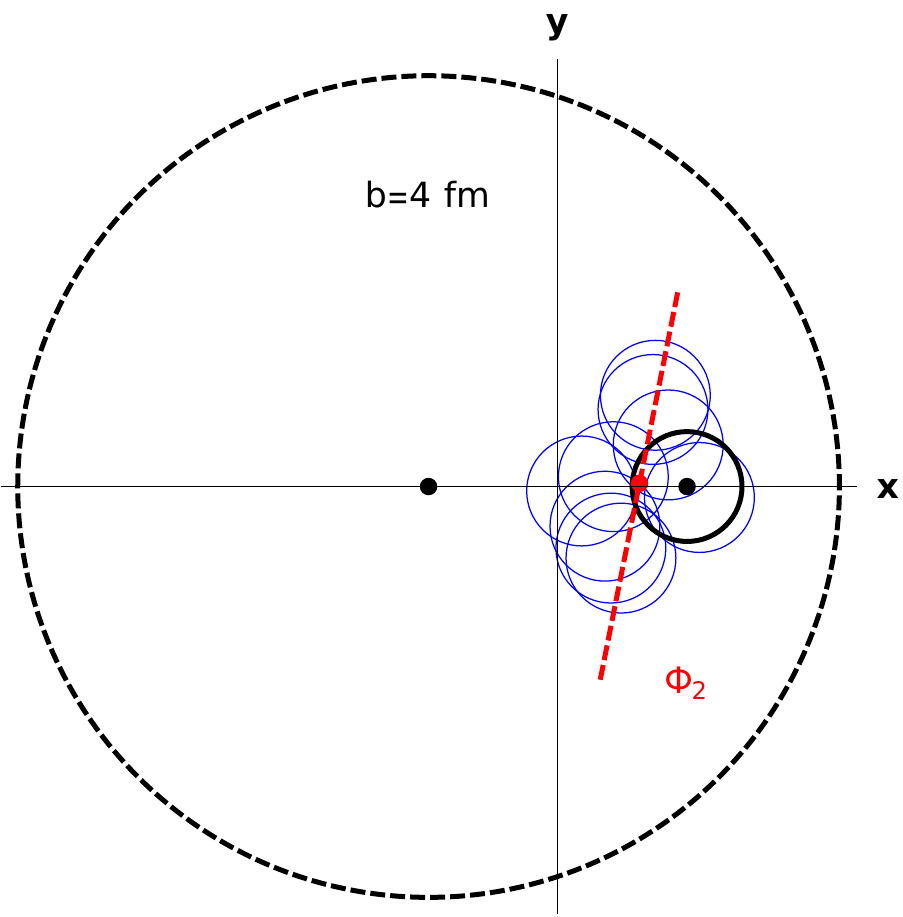}
			\includegraphics[scale=0.7]{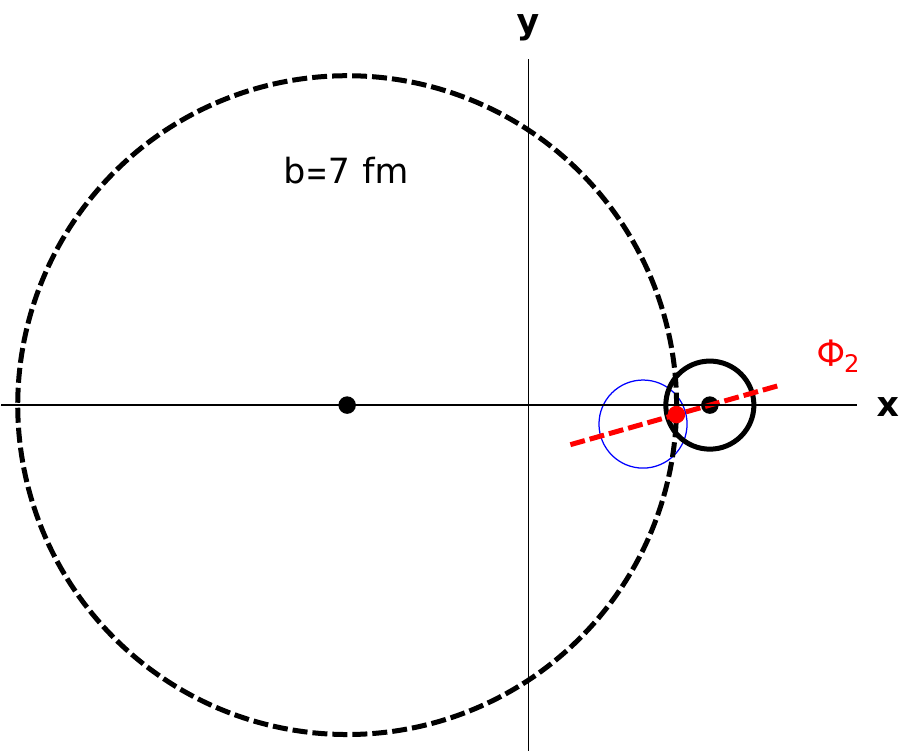}
	\caption{	\label{fig:participants_pA} Examples of participants distribution in one event with $b=4$ fm (left panel), and $b=7$ fm (right panel). The red point is centre of mass $\bf{r}_c$, and the the red line indicates $\Phi_2$. The blue circles are partipants from target nucleus, while the black circle is the single projectile proton.}
\end{figure}

Shown on \fig{fig:fields-pA-HS-By}, total magnetic field $B_y$ contributed mainly from spectators. The strength of total $B_y$ and the strength from spectators are both increased with impact parameter, and have maximal values at $b \approx 6.5$fm.
While in \fig{fig:fields-pA-HS-Ex}, contribution from participants to $E_x$ is negative in the range of impact parameter $b$  between 2 and 8 fm. This is simply because that, the single projectile proton collisions on periphery of target nucleus in this range of $b$. So the center of mass $\bf{r_c}$  is approximately geometry centre of all participants in each event, the single projectile proton prefer locating on $+x$ side of geometry centre, as shown in the illustration  \fig{fig:participants_pA}.
After average over many events, contributions from target participants to $E_x$ will offset with each other, but contribution from single projectile proton will remain along $-x$ direction.

\begin{figure}[htbp]
\centering
	\subfigure{
		\label{fig:fields-pA-CP-Ex}
		\includegraphics[scale=0.6]{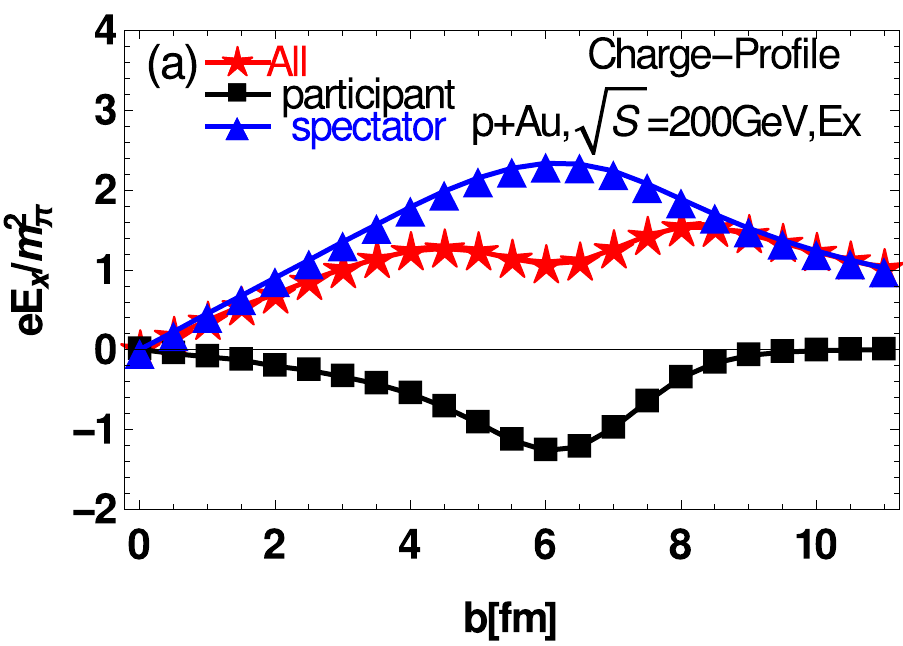}}
	\subfigure{
		\label{fig:fields-pA-CP-By}
		\includegraphics[scale=0.6]{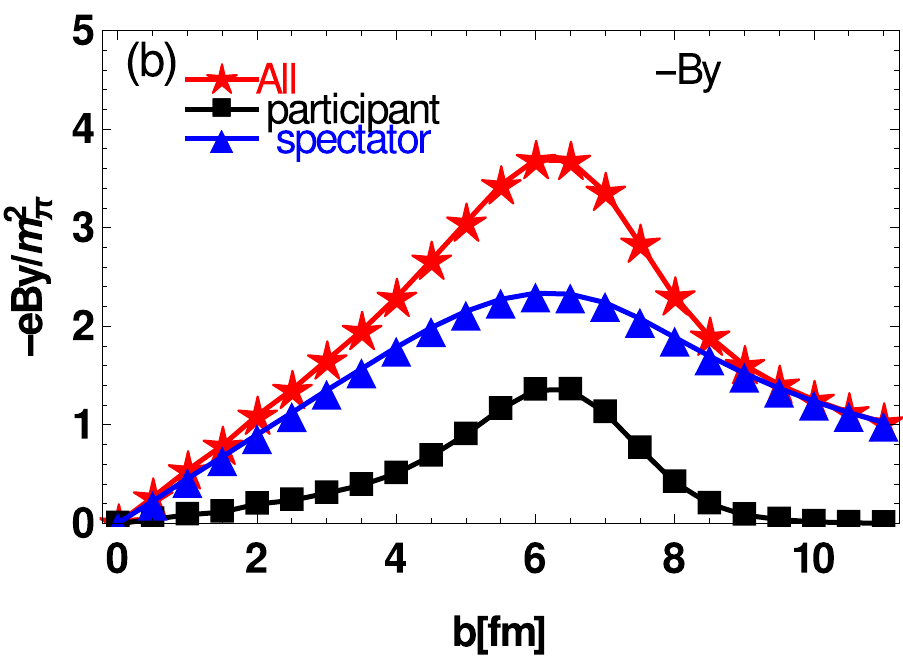}}
	\subfigure{
		\label{fig:fields-pA-CP-absBx}
		\includegraphics[scale=0.6]{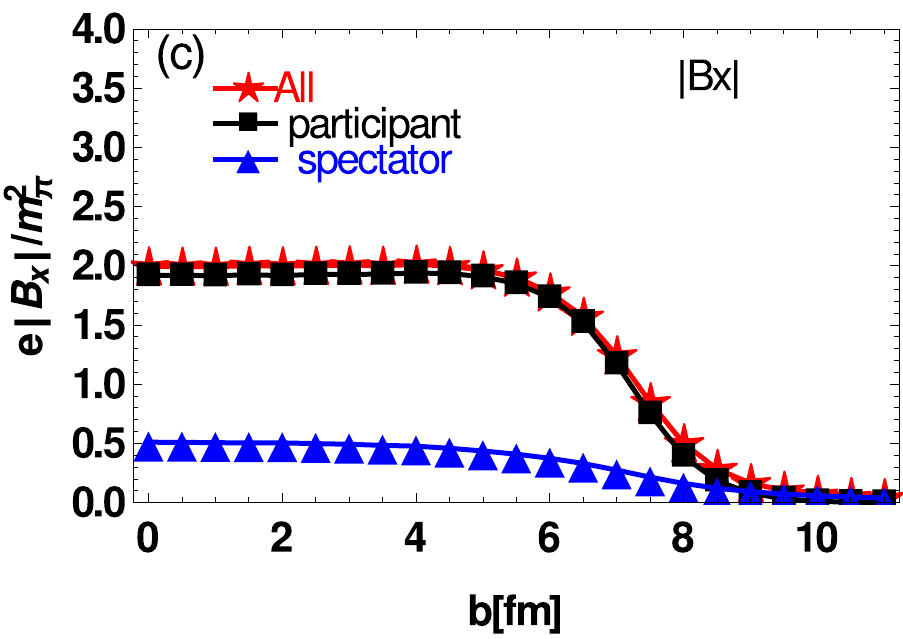}}
	\subfigure{
		\label{fig:fields-pA-CP-absBy}
		\includegraphics[scale=0.6]{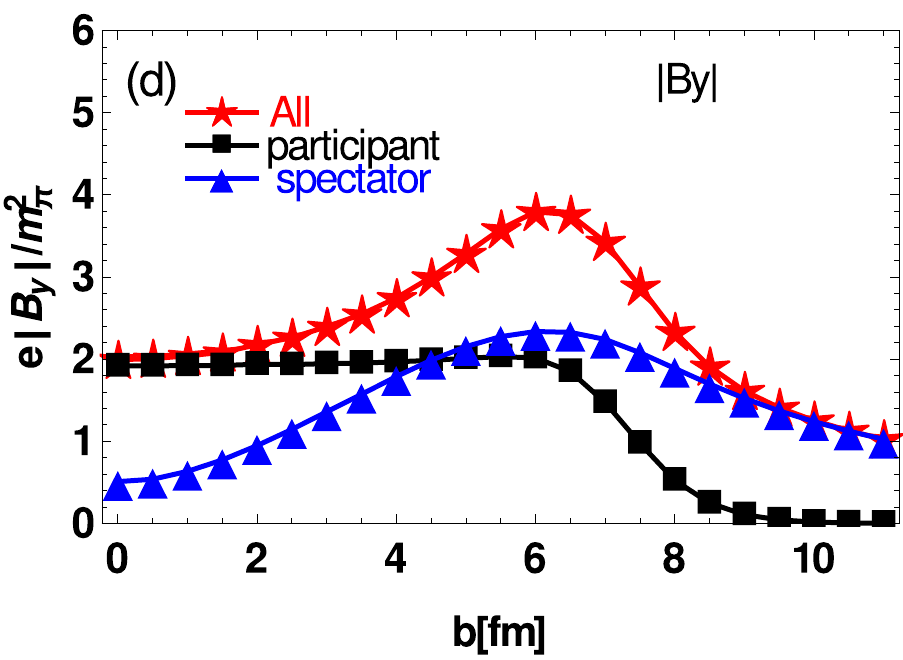}}
\caption{\label{fig:fields-pA-CP}The electromagnetic fields at $t$=0 and $\bf{r}=\bf{r}_c$ as function of the impact parameter b with Charge-Profile model.}
\end{figure}

In \fig{fig:fields-pA-CP}, results of impact parameter dependence of fields calculated with Charge-Profile model are given. Since the 3-dimensional charge density are given from electromagnetic form factors measured in experiments\cite{Alberico:2008sz}\cite{Miller:2010nz}, these results are more physical than those with Hard-Sphere model.

Firstly, let's check the fluctuations with Charge-Profile model. Shown in \fig{fig:fields-pA-CP-absBx} and \fig{fig:fields-pA-CP-absBy}, we can see fluctuation of fields from spectators are almost identical with that in Hard-Sphere model. That means the contribution from spectators are not sensitive to the detail of inner charge distribution of nucleon. This is because spectators are far away from $\bf{r}_c$, the discrepancy of inner charge density details can not be seen clearly. On the contrary, since all participants are near to $\bf{r}_c$, their contributions are not negligible, so the fluctuations from participants are quite different with that in Hard-Sphere model.

About the strength of electric fields $E_x$ shown on \fig{fig:fields-pA-CP-Ex}, contribution from all participant is still negative, like that in Hard-Sphere model, since the contribution from single projectile proton remains along $-x$ direction. Its negative strength is so  large that there is a significant depression to total $E_x$ at $b \approx 6$ fm.

In \fig{fig:fields-pA-CP-By}, although strength of $B_y$ contributed from spectators are still similar with Hard-Sphere model, the contribution from participants is quite larger. Especially in peripheral collisions, contributions from participants are even with same order with spectators. This is the main difference comparing with heavy ion $A+A$ collision system.  Since the orientation of event plane is determined only by participants, this large contribution of participants to $B_y$ maybe bring a significant azimuthal correlation between event plane $\Phi_2$ and magnetic field $\Phi_B$.

Comparing the magnetic strength of $B_y$ in paper \cite{Belmont:2016oqp}, our results within Charge-Profile model is about half. Consiering the very small size of overlapping area in small collision system, any discrepency in the inner charge density of nucleion could lead to a significant difference on fields strength. This could also be concluded while comparing \fig{fig:field-singleP-z0} and \fig{fig:field-singleP-PL-z0}.  So, beyond the Hard-Sphere model, we also implement the realistic Charge-Profile model in our calculation.

\subsection{Azimuthal correlation between $\bf{B}$ and reaction plane}
As we mentioned, $\Delta\gamma\propto B^2\cdot\cos2(\Phi_B-\Phi_2)$ , where $\Phi_B$ is the angle of magnetic field  $\bf{B}$ direction. In this section we focus on the 2nd harmonic participant plane $\Phi_2$ as it is the most prominent anisotropy from both geometry and fluctuations.
The corresponding initial event plane angles are given by:\\
\begin{equation}
 \Phi_\emph{n}=\frac{1}{\emph{n}}\arctan\frac{<\emph{r}^\emph{n}\sin(\emph{n}\phi)>}{<\emph{r}^\emph{n}\cos(\emph{n}\phi)>}
\end{equation}
where $\emph{r}$ is the distance of the participants particles from the center of mass $\bf{r}_c$, $\phi$ is  angle between the x direction and the direction of $\emph{r} $~\cite{Deng:2011at}. We define $\Phi_2$ as the long axis direction of the ellipse.

\begin{figure}[htbp]
	\centering
		\includegraphics[scale=0.6]{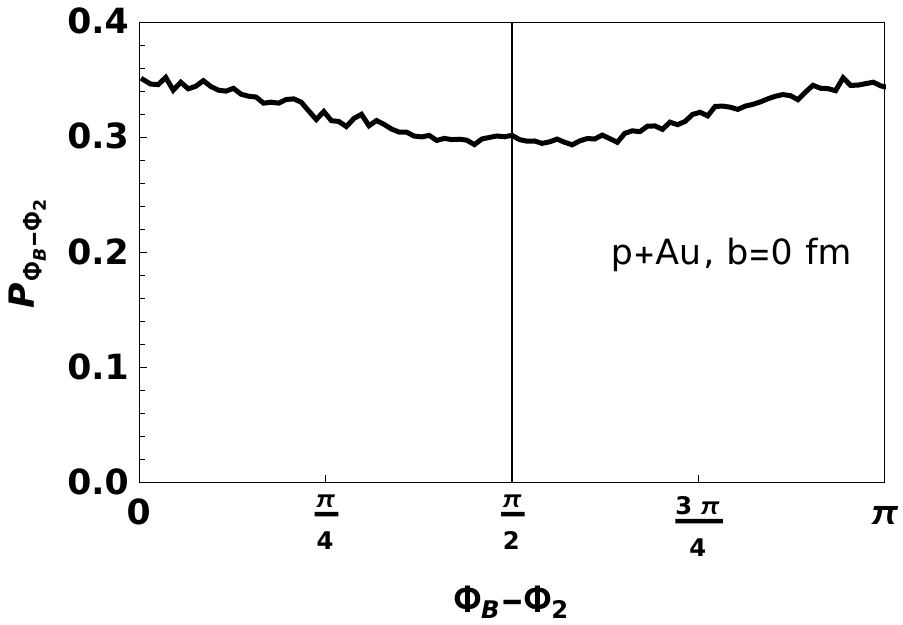}
			\includegraphics[scale=0.6]{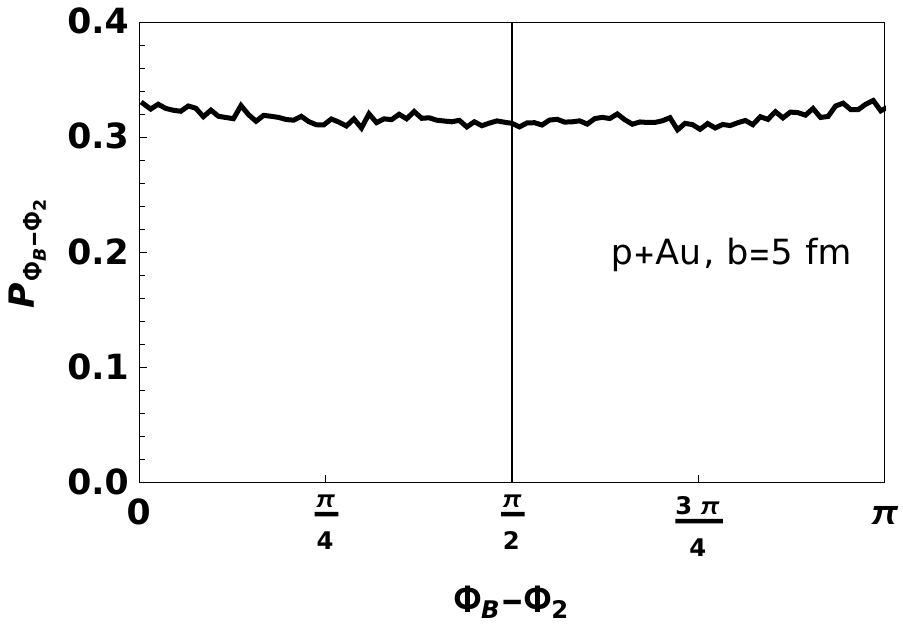}
		\includegraphics[scale=0.6]{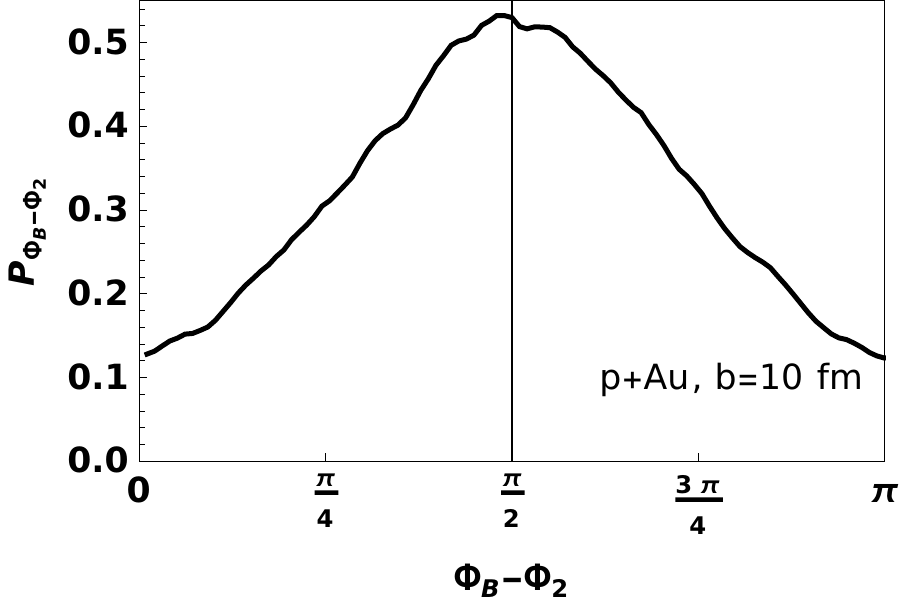}
	\caption{\label{fig:P_angle_correlation_RHIC}The electromagnetic fields at $t$=0 and $\bf{r}=\bf{r}_c$ as function of the impact parameter b with Charge-Profile model.}
\end{figure}

Our calculations in this subsection are all within Charge-Profile model to get a physical understanding of azimuthal correlation between $\Phi_B$ and  $\Phi_2$. Firstly, we give the distribution of the discrepancy $\Phi_B - \Phi_2$ in \fig{fig:P_angle_correlation_RHIC}.
We can see there is a strong negative correlation between $\Phi_B$ and $\Phi_2$ observed at large $b$, this is because there could be only one nucleon in target nucleus hit by single projectile proton in this very peripheral collision, as illustrated in right panel of \fig{fig:participants_pA}. So the direction of $\Phi_2$ (the long axis of ellipse) prefers near to $x$ direction, which is perpendicular to the direction of magnetic field.

While at small  and middle values of $b$, there is a slight but not negligible correlation between $\Phi_B$ and $\Phi_2$, even at $b=0$ fm shown in the first panel on \fig{fig:P_angle_correlation_RHIC}. This phenomenon could also be explained from left panel in \fig{fig:participants_pA}. In each event, the contribution to fields from the single projectile proton is very large comparing to target participants since the centre of mass $\bf{r}_c$ is very near to it. While the location of this single projectile proton prefers to $+x$ side of $\bf{r}_c$ in event, its contribution to $B_y$ prefers to $-y$ direction in each event. On the other hand, due to nuclear geometry, the target participants overlap as an approximate ellipse with its long axis along $y$ direction.

\begin{figure}[htbp]
\centering
\includegraphics[scale=0.5]{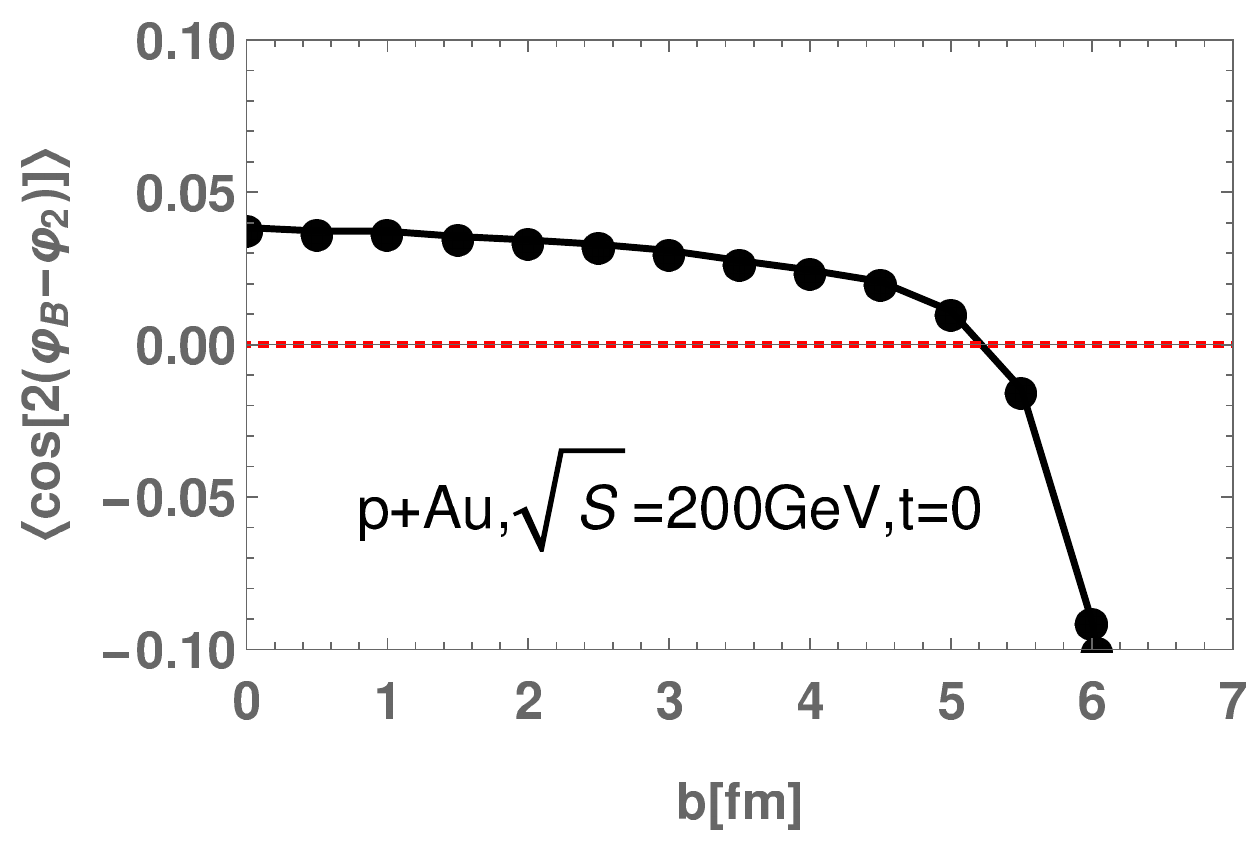}
\includegraphics[scale=0.5]{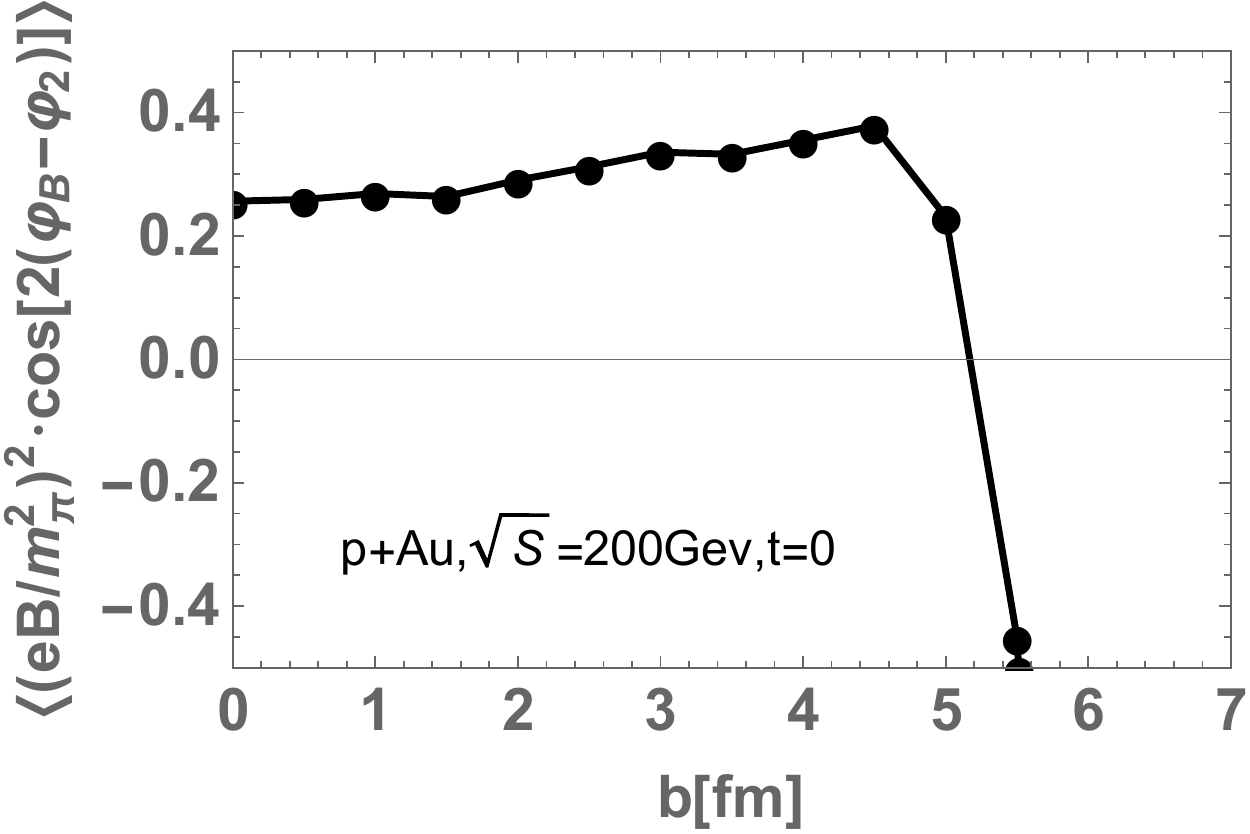}
\caption{\label{fig:azimuthal-correlation} The left panel is the azimuthal correlations between magnetic field plane $\Phi_B$ and $\Phi_2$, while the right panel is $B^2\cdot\cos2(\Phi_B-\Phi_n)$ as a function of impact parameter for $p+Au$ at RHIC energy .}
\end{figure}

This correlation between $\Phi_B$ and $\Phi_2$ could also shown in the azimuthal correlation factor $\cos2(\Phi_B-\Phi_2)$, shown in left panel of \fig{fig:azimuthal-correlation}. We can see the value of this correlation factor is very small but not vanished at small and middle range of impact parameter $b$. In large $b$ range, the azimuthal correlation shown in the lower panel of \fig{fig:P_angle_correlation_RHIC} appears again here. This result is also consistant with paper\cite{Zhao_2018}.

Since the CME induced charge separation effect is described as $\Delta\gamma\propto B^2\cdot\cos2(\Phi_B-\Phi_2)$, We give this $B^2$ evolved azimuthal correlation in right panel of \fig{fig:azimuthal-correlation}.
Due to the large fluctuation of strength of magnetic field (\fig{fig:fields-pA-CP-absBy}), the slight remaining  azimuthal correlation could give a finite  $\Delta\gamma$ signal due to the CME effect.

\begin{figure}[htbp]
	\centering
	\includegraphics[scale=0.5]{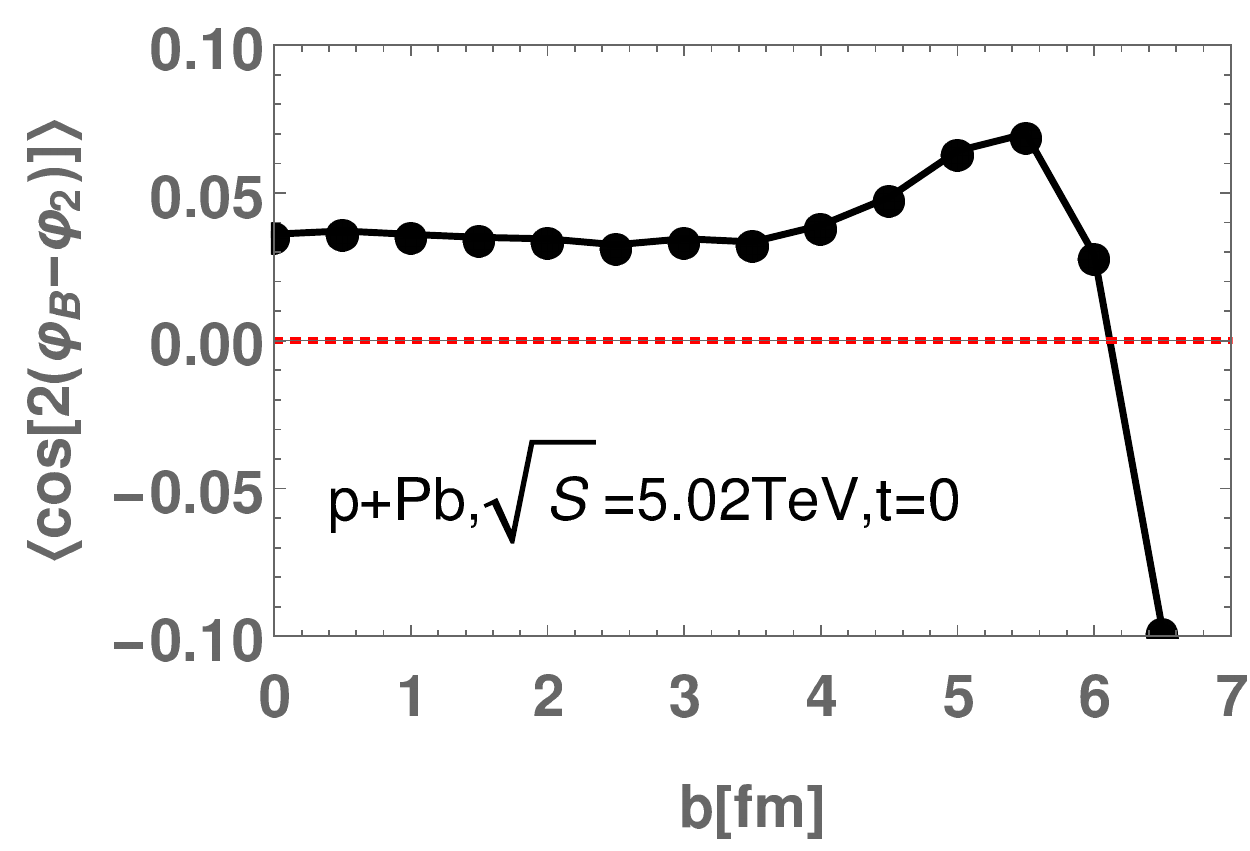}
	\includegraphics[scale=0.5]{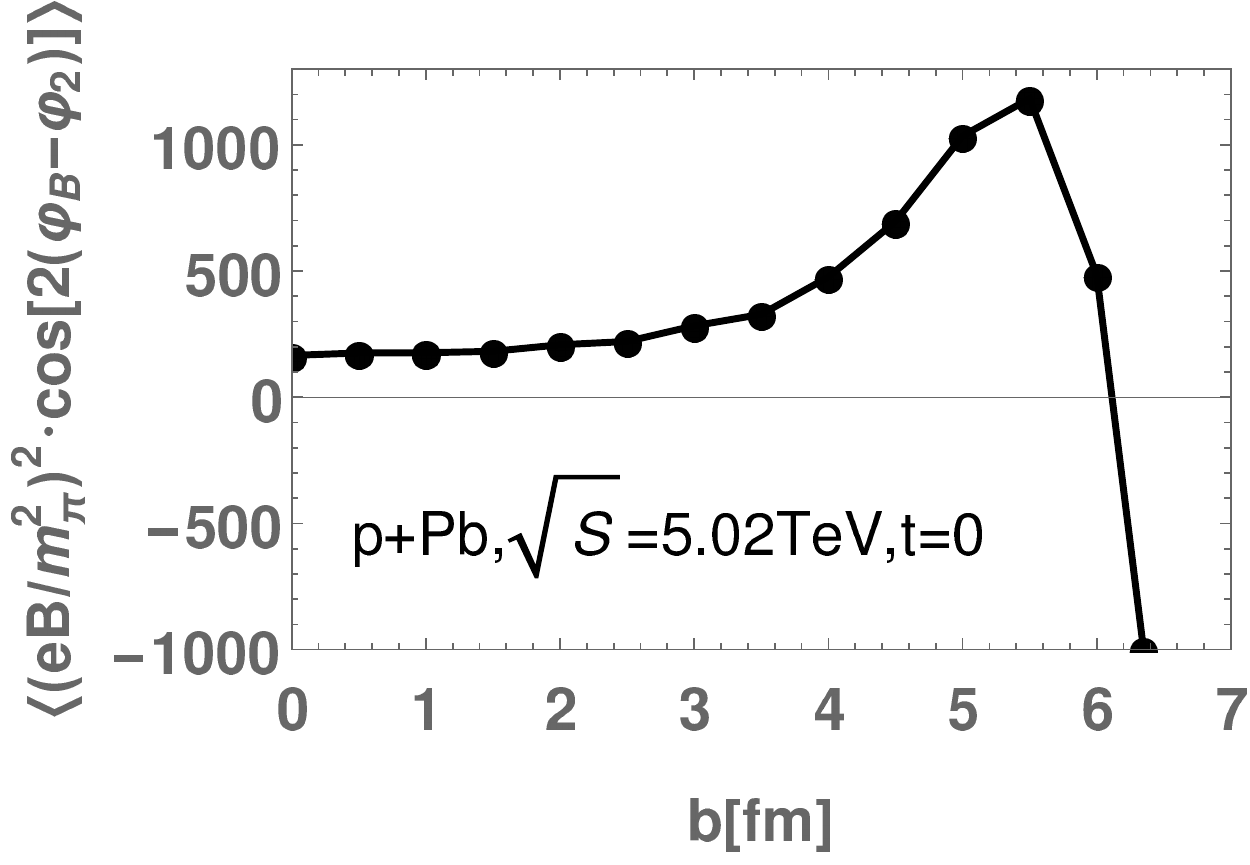}
	\caption{\label{fig:azimuthal-correlation_5.02TeV} The left panel is the azimuthal correlations between magnetic field plane $\Phi_B$ and $\Phi_2$, while the right panel is $B^2\cdot\cos2(\Phi_B-\Phi_n)$ as a function of impact parameter for $p+Pb$ in LHC energy .}
\end{figure}

This calculation can be extended to LHC energy easily. We give corresponding results in \fig{fig:azimuthal-correlation_5.02TeV}. The result of azimuthal correlation $\cos2(\Phi_B-\Phi_2)$ at LHC energy is similar with what at RHIC energy. While due to much more fluctuation of the strength of magnetic field, the $B^2$ evolved azimuthal correlation $B^2\cdot\cos2(\Phi_B-\Phi_n)$ in LHC energy is three order larger than what in RHIC energy.
With this strong azimuthal correlation between $\Phi_B$ and $\Phi_2$, it implies that the CME signals may be found in the small system, and maybe explain the observed $\Delta\gamma$ experiment data partly \cite{Sirunyan:2017quh,Khachatryan:2016got}.

\section{CONCLUSION}

The contribution of CME to the charge separation effect related not only with the strength of magnetic field, but also with the angle between direction of magnetic field $\Phi_B$ and reaction plane $\Phi_2$. In the heavy ion small collision system with p+A or d+A collision, people used to believe that the direction of magnetic field is determined by the distribution of spectators mainly, while the direction of reaction plane is computed from the space distribution of participants. Due to the decoupling of angular correlation between the reaction plan and direction of magnetic field, this charge separation effect should be vanished. However, the experiment data show a quite different result comparing to theory predictions.

Employing a physical Charge-Profile model to describe the inner charge distribution of a proton and a neutron, we calculated the property of electromagnetic field produced in small system both in RHIC and LHC energy systematically, including its dependence on impact parameter b. Especially, we have studied the azimuthal correlation  between $\Phi_B$ and $\Phi_2$ carefully.

In conflict with people's previous expectations, our results show a significant angular correlation between them. This correlation comes mainly from the very close distance between the location of single projectile proton and the location of C.M.S $\bf{r}_c$ of hot medium produced in small collision system. Then the contribution of single projectile proton to the magnetic field is the main source after average over all participants. So the strong angular distribution priority of this single projectile proton could contribute significantly to the angular correlation of $\Phi_B$ and $\Phi_2$.

This discovery breakthrough people's previous thought about magnetic filed produced in small collision system. It can help us to clarify contribution of CME to the observation in small system collision experiments.

\section{Acknowledgments:}
We thank Xu-Guang Huang for very helpful discussions. We were supported by the NSFC under Projects No. 12075094 and No. 11535005. 

%%%%%%%%%%%%%%%%%%%%%%%%%%%%%%%%%%%%%%%%%%%%%%%%%%%%%%%%%%%%%%%%%%%
\bibliographystyle{apsrev4-2}
\bibliography{reference}

\begin{thebibliography}{35}%
\makeatletter
\providecommand \@ifxundefined [1]{%
 \@ifx{#1\undefined}
}%
\providecommand \@ifnum [1]{%
 \ifnum #1\expandafter \@firstoftwo
 \else \expandafter \@secondoftwo
 \fi
}%
\providecommand \@ifx [1]{%
 \ifx #1\expandafter \@firstoftwo
 \else \expandafter \@secondoftwo
 \fi
}%
\providecommand \natexlab [1]{#1}%
\providecommand \enquote  [1]{``#1''}%
\providecommand \bibnamefont  [1]{#1}%
\providecommand \bibfnamefont [1]{#1}%
\providecommand \citenamefont [1]{#1}%
\providecommand \href@noop [0]{\@secondoftwo}%
\providecommand \href [0]{\begingroup \@sanitize@url \@href}%
\providecommand \@href[1]{\@@startlink{#1}\@@href}%
\providecommand \@@href[1]{\endgroup#1\@@endlink}%
\providecommand \@sanitize@url [0]{\catcode `\\12\catcode `\$12\catcode
  `\&12\catcode `\#12\catcode `\^12\catcode `\_12\catcode `\%12\relax}%
\providecommand \@@startlink[1]{}%
\providecommand \@@endlink[0]{}%
\providecommand \url  [0]{\begingroup\@sanitize@url \@url }%
\providecommand \@url [1]{\endgroup\@href {#1}{\urlprefix }}%
\providecommand \urlprefix  [0]{URL }%
\providecommand \Eprint [0]{\href }%
\providecommand \doibase [0]{https://doi.org/}%
\providecommand \selectlanguage [0]{\@gobble}%
\providecommand \bibinfo  [0]{\@secondoftwo}%
\providecommand \bibfield  [0]{\@secondoftwo}%
\providecommand \translation [1]{[#1]}%
\providecommand \BibitemOpen [0]{}%
\providecommand \bibitemStop [0]{}%
\providecommand \bibitemNoStop [0]{.\EOS\space}%
\providecommand \EOS [0]{\spacefactor3000\relax}%
\providecommand \BibitemShut  [1]{\csname bibitem#1\endcsname}%
\let\auto@bib@innerbib\@empty
%</preamble>
\bibitem [{\citenamefont {Rafelski}\ and\ \citenamefont
  {Muller}(1976)}]{Rafelski:1975rf}%
  \BibitemOpen
  \bibfield  {author} {\bibinfo {author} {\bibfnamefont {J.}~\bibnamefont
  {Rafelski}}\ and\ \bibinfo {author} {\bibfnamefont {B.}~\bibnamefont
  {Muller}},\ }\href {https://doi.org/10.1103/PhysRevLett.36.517} {\bibfield
  {journal} {\bibinfo  {journal} {Phys. Rev. Lett.}\ }\textbf {\bibinfo
  {volume} {36}},\ \bibinfo {pages} {517} (\bibinfo {year} {1976})}\BibitemShut
  {NoStop}%
%%CITATION = PRLTA,36,517;%%
\bibitem [{\citenamefont {Skokov}\ \emph {et~al.}(2009)\citenamefont {Skokov},
  \citenamefont {Illarionov},\ and\ \citenamefont {Toneev}}]{Skokov:2009qp}%
  \BibitemOpen
  \bibfield  {author} {\bibinfo {author} {\bibfnamefont {V.}~\bibnamefont
  {Skokov}}, \bibinfo {author} {\bibfnamefont {A.~{\relax Yu}.}\ \bibnamefont
  {Illarionov}},\ and\ \bibinfo {author} {\bibfnamefont {V.}~\bibnamefont
  {Toneev}},\ }\href {https://doi.org/10.1142/S0217751X09047570} {\bibfield
  {journal} {\bibinfo  {journal} {Int. J. Mod. Phys.}\ }\textbf {\bibinfo
  {volume} {A24}},\ \bibinfo {pages} {5925} (\bibinfo {year} {2009})},\ \Eprint
  {https://arxiv.org/abs/0907.1396} {arXiv:0907.1396 [nucl-th]} \BibitemShut
  {NoStop}%
%%CITATION = ARXIV:0907.1396;%%
\bibitem [{\citenamefont {Bzdak}\ and\ \citenamefont
  {Skokov}(2012)}]{Bzdak:2011yy}%
  \BibitemOpen
  \bibfield  {author} {\bibinfo {author} {\bibfnamefont {A.}~\bibnamefont
  {Bzdak}}\ and\ \bibinfo {author} {\bibfnamefont {V.}~\bibnamefont {Skokov}},\
  }\href {https://doi.org/10.1016/j.physletb.2012.02.065} {\bibfield  {journal}
  {\bibinfo  {journal} {Phys. Lett. B}\ }\textbf {\bibinfo {volume} {710}},\
  \bibinfo {pages} {171} (\bibinfo {year} {2012})},\ \Eprint
  {https://arxiv.org/abs/1111.1949} {arXiv:1111.1949 [hep-ph]} \BibitemShut
  {NoStop}%
\bibitem [{\citenamefont {Voronyuk}\ \emph {et~al.}(2011)\citenamefont
  {Voronyuk}, \citenamefont {Toneev}, \citenamefont {Cassing}, \citenamefont
  {Bratkovskaya}, \citenamefont {Konchakovski},\ and\ \citenamefont
  {Voloshin}}]{Voronyuk:2011jd}%
  \BibitemOpen
  \bibfield  {author} {\bibinfo {author} {\bibfnamefont {V.}~\bibnamefont
  {Voronyuk}}, \bibinfo {author} {\bibfnamefont {V.~D.}\ \bibnamefont
  {Toneev}}, \bibinfo {author} {\bibfnamefont {W.}~\bibnamefont {Cassing}},
  \bibinfo {author} {\bibfnamefont {E.~L.}\ \bibnamefont {Bratkovskaya}},
  \bibinfo {author} {\bibfnamefont {V.~P.}\ \bibnamefont {Konchakovski}},\ and\
  \bibinfo {author} {\bibfnamefont {S.~A.}\ \bibnamefont {Voloshin}},\ }\href
  {https://doi.org/10.1103/PhysRevC.83.054911} {\bibfield  {journal} {\bibinfo
  {journal} {Phys. Rev. C}\ }\textbf {\bibinfo {volume} {83}},\ \bibinfo
  {pages} {054911} (\bibinfo {year} {2011})},\ \Eprint
  {https://arxiv.org/abs/1103.4239} {arXiv:1103.4239 [nucl-th]} \BibitemShut
  {NoStop}%
\bibitem [{\citenamefont {Deng}\ and\ \citenamefont
  {Huang}(2012)}]{Deng:2012pc}%
  \BibitemOpen
  \bibfield  {author} {\bibinfo {author} {\bibfnamefont {W.-T.}\ \bibnamefont
  {Deng}}\ and\ \bibinfo {author} {\bibfnamefont {X.-G.}\ \bibnamefont
  {Huang}},\ }\href {https://doi.org/10.1103/PhysRevC.85.044907} {\bibfield
  {journal} {\bibinfo  {journal} {Phys. Rev.}\ }\textbf {\bibinfo {volume}
  {C85}},\ \bibinfo {pages} {044907} (\bibinfo {year} {2012})},\ \Eprint
  {https://arxiv.org/abs/1201.5108} {arXiv:1201.5108 [nucl-th]} \BibitemShut
  {NoStop}%
%%CITATION = ARXIV:1201.5108;%%
\bibitem [{\citenamefont {Deng}\ and\ \citenamefont
  {Huang}(2015)}]{Deng:2014uja}%
  \BibitemOpen
  \bibfield  {author} {\bibinfo {author} {\bibfnamefont {W.-T.}\ \bibnamefont
  {Deng}}\ and\ \bibinfo {author} {\bibfnamefont {X.-G.}\ \bibnamefont
  {Huang}},\ }\href {https://doi.org/10.1016/j.physletb.2015.01.050} {\bibfield
   {journal} {\bibinfo  {journal} {Phys. Lett. B}\ }\textbf {\bibinfo {volume}
  {742}},\ \bibinfo {pages} {296} (\bibinfo {year} {2015})},\ \Eprint
  {https://arxiv.org/abs/1411.2733} {arXiv:1411.2733 [nucl-th]} \BibitemShut
  {NoStop}%
\bibitem [{\citenamefont {Inghirami}\ \emph {et~al.}(2016)\citenamefont
  {Inghirami}, \citenamefont {Del~Zanna}, \citenamefont {Beraudo},
  \citenamefont {Moghaddam}, \citenamefont {Becattini},\ and\ \citenamefont
  {Bleicher}}]{Inghirami:2016iru}%
  \BibitemOpen
  \bibfield  {author} {\bibinfo {author} {\bibfnamefont {G.}~\bibnamefont
  {Inghirami}}, \bibinfo {author} {\bibfnamefont {L.}~\bibnamefont
  {Del~Zanna}}, \bibinfo {author} {\bibfnamefont {A.}~\bibnamefont {Beraudo}},
  \bibinfo {author} {\bibfnamefont {M.~H.}\ \bibnamefont {Moghaddam}}, \bibinfo
  {author} {\bibfnamefont {F.}~\bibnamefont {Becattini}},\ and\ \bibinfo
  {author} {\bibfnamefont {M.}~\bibnamefont {Bleicher}},\ }\href
  {https://doi.org/10.1140/epjc/s10052-016-4516-8} {\bibfield  {journal}
  {\bibinfo  {journal} {Eur. Phys. J. C}\ }\textbf {\bibinfo {volume} {76}},\
  \bibinfo {pages} {659} (\bibinfo {year} {2016})},\ \Eprint
  {https://arxiv.org/abs/1609.03042} {arXiv:1609.03042 [hep-ph]} \BibitemShut
  {NoStop}%
\bibitem [{\citenamefont {Yan}\ and\ \citenamefont
  {Huang}(2021)}]{Yan:2021zjc}%
  \BibitemOpen
  \bibfield  {author} {\bibinfo {author} {\bibfnamefont {L.}~\bibnamefont
  {Yan}}\ and\ \bibinfo {author} {\bibfnamefont {X.-G.}\ \bibnamefont
  {Huang}},\ }\href@noop {} {\  (\bibinfo {year} {2021})},\ \Eprint
  {https://arxiv.org/abs/2104.00831} {arXiv:2104.00831 [nucl-th]} \BibitemShut
  {NoStop}%
\bibitem [{\citenamefont {Kharzeev}\ \emph {et~al.}(1998)\citenamefont
  {Kharzeev}, \citenamefont {Pisarski},\ and\ \citenamefont
  {Tytgat}}]{Kharzeev:1998kz}%
  \BibitemOpen
  \bibfield  {author} {\bibinfo {author} {\bibfnamefont {D.}~\bibnamefont
  {Kharzeev}}, \bibinfo {author} {\bibfnamefont {R.~D.}\ \bibnamefont
  {Pisarski}},\ and\ \bibinfo {author} {\bibfnamefont {M.~H.~G.}\ \bibnamefont
  {Tytgat}},\ }\href {https://doi.org/10.1103/PhysRevLett.81.512} {\bibfield
  {journal} {\bibinfo  {journal} {Phys. Rev. Lett.}\ }\textbf {\bibinfo
  {volume} {81}},\ \bibinfo {pages} {512} (\bibinfo {year} {1998})},\ \Eprint
  {https://arxiv.org/abs/hep-ph/9804221} {arXiv:hep-ph/9804221 [hep-ph]}
  \BibitemShut {NoStop}%
%%CITATION = HEP-PH/9804221;%%
\bibitem [{\citenamefont {Kharzeev}(2006)}]{Kharzeev:2004ey}%
  \BibitemOpen
  \bibfield  {author} {\bibinfo {author} {\bibfnamefont {D.}~\bibnamefont
  {Kharzeev}},\ }\href {https://doi.org/10.1016/j.physletb.2005.11.075}
  {\bibfield  {journal} {\bibinfo  {journal} {Phys. Lett.}\ }\textbf {\bibinfo
  {volume} {B633}},\ \bibinfo {pages} {260} (\bibinfo {year} {2006})},\ \Eprint
  {https://arxiv.org/abs/hep-ph/0406125} {arXiv:hep-ph/0406125 [hep-ph]}
  \BibitemShut {NoStop}%
%%CITATION = HEP-PH/0406125;%%
\bibitem [{\citenamefont {Kharzeev}\ \emph {et~al.}(2008)\citenamefont
  {Kharzeev}, \citenamefont {McLerran},\ and\ \citenamefont
  {Warringa}}]{Kharzeev:2007jp}%
  \BibitemOpen
  \bibfield  {author} {\bibinfo {author} {\bibfnamefont {D.~E.}\ \bibnamefont
  {Kharzeev}}, \bibinfo {author} {\bibfnamefont {L.~D.}\ \bibnamefont
  {McLerran}},\ and\ \bibinfo {author} {\bibfnamefont {H.~J.}\ \bibnamefont
  {Warringa}},\ }\href {https://doi.org/10.1016/j.nuclphysa.2008.02.298}
  {\bibfield  {journal} {\bibinfo  {journal} {Nucl. Phys.}\ }\textbf {\bibinfo
  {volume} {A803}},\ \bibinfo {pages} {227} (\bibinfo {year} {2008})},\ \Eprint
  {https://arxiv.org/abs/0711.0950} {arXiv:0711.0950 [hep-ph]} \BibitemShut
  {NoStop}%
%%CITATION = ARXIV:0711.0950;%%
\bibitem [{\citenamefont {Fukushima}\ \emph {et~al.}(2008)\citenamefont
  {Fukushima}, \citenamefont {Kharzeev},\ and\ \citenamefont
  {Warringa}}]{Fukushima:2008xe}%
  \BibitemOpen
  \bibfield  {author} {\bibinfo {author} {\bibfnamefont {K.}~\bibnamefont
  {Fukushima}}, \bibinfo {author} {\bibfnamefont {D.~E.}\ \bibnamefont
  {Kharzeev}},\ and\ \bibinfo {author} {\bibfnamefont {H.~J.}\ \bibnamefont
  {Warringa}},\ }\href {https://doi.org/10.1103/PhysRevD.78.074033} {\bibfield
  {journal} {\bibinfo  {journal} {Phys. Rev.}\ }\textbf {\bibinfo {volume}
  {D78}},\ \bibinfo {pages} {074033} (\bibinfo {year} {2008})},\ \Eprint
  {https://arxiv.org/abs/0808.3382} {arXiv:0808.3382 [hep-ph]} \BibitemShut
  {NoStop}%
%%CITATION = ARXIV:0808.3382;%%
\bibitem [{\citenamefont {Kharzeev}\ \emph {et~al.}(2016)\citenamefont
  {Kharzeev}, \citenamefont {Liao}, \citenamefont {Voloshin},\ and\
  \citenamefont {Wang}}]{Kharzeev:2015znc}%
  \BibitemOpen
  \bibfield  {author} {\bibinfo {author} {\bibfnamefont {D.~E.}\ \bibnamefont
  {Kharzeev}}, \bibinfo {author} {\bibfnamefont {J.}~\bibnamefont {Liao}},
  \bibinfo {author} {\bibfnamefont {S.~A.}\ \bibnamefont {Voloshin}},\ and\
  \bibinfo {author} {\bibfnamefont {G.}~\bibnamefont {Wang}},\ }\href
  {https://doi.org/10.1016/j.ppnp.2016.01.001} {\bibfield  {journal} {\bibinfo
  {journal} {Prog. Part. Nucl. Phys.}\ }\textbf {\bibinfo {volume} {88}},\
  \bibinfo {pages} {1} (\bibinfo {year} {2016})},\ \Eprint
  {https://arxiv.org/abs/1511.04050} {arXiv:1511.04050 [hep-ph]} \BibitemShut
  {NoStop}%
%%CITATION = ARXIV:1511.04050;%%
\bibitem [{\citenamefont {Liu}\ and\ \citenamefont
  {Huang}(2020)}]{Liu:2020ymh}%
  \BibitemOpen
  \bibfield  {author} {\bibinfo {author} {\bibfnamefont {Y.-C.}\ \bibnamefont
  {Liu}}\ and\ \bibinfo {author} {\bibfnamefont {X.-G.}\ \bibnamefont
  {Huang}},\ }\href {https://doi.org/10.1007/s41365-020-00764-z} {\bibfield
  {journal} {\bibinfo  {journal} {Nucl. Sci. Tech.}\ }\textbf {\bibinfo
  {volume} {31}},\ \bibinfo {pages} {56} (\bibinfo {year} {2020})},\ \Eprint
  {https://arxiv.org/abs/2003.12482} {arXiv:2003.12482 [nucl-th]} \BibitemShut
  {NoStop}%
\bibitem [{\citenamefont {Voloshin}(2004)}]{Voloshin:2004vk}%
  \BibitemOpen
  \bibfield  {author} {\bibinfo {author} {\bibfnamefont {S.~A.}\ \bibnamefont
  {Voloshin}},\ }\href {https://doi.org/10.1103/PhysRevC.70.057901} {\bibfield
  {journal} {\bibinfo  {journal} {Phys. Rev.}\ }\textbf {\bibinfo {volume}
  {C70}},\ \bibinfo {pages} {057901} (\bibinfo {year} {2004})},\ \Eprint
  {https://arxiv.org/abs/hep-ph/0406311} {arXiv:hep-ph/0406311 [hep-ph]}
  \BibitemShut {NoStop}%
%%CITATION = HEP-PH/0406311;%%
\bibitem [{\citenamefont {Abelev}\ \emph {et~al.}(2009)\citenamefont {Abelev}
  \emph {et~al.}}]{Abelev:2009ac}%
  \BibitemOpen
  \bibfield  {author} {\bibinfo {author} {\bibfnamefont {B.~I.}\ \bibnamefont
  {Abelev}} \emph {et~al.} (\bibinfo {collaboration} {STAR}),\ }\href
  {https://doi.org/10.1103/PhysRevLett.103.251601} {\bibfield  {journal}
  {\bibinfo  {journal} {Phys. Rev. Lett.}\ }\textbf {\bibinfo {volume} {103}},\
  \bibinfo {pages} {251601} (\bibinfo {year} {2009})},\ \Eprint
  {https://arxiv.org/abs/0909.1739} {arXiv:0909.1739 [nucl-ex]} \BibitemShut
  {NoStop}%
%%CITATION = ARXIV:0909.1739;%%
\bibitem [{\citenamefont {Abelev}\ \emph {et~al.}(2010)\citenamefont {Abelev}
  \emph {et~al.}}]{Abelev:2009ad}%
  \BibitemOpen
  \bibfield  {author} {\bibinfo {author} {\bibfnamefont {B.~I.}\ \bibnamefont
  {Abelev}} \emph {et~al.} (\bibinfo {collaboration} {STAR}),\ }\href
  {https://doi.org/10.1103/PhysRevC.81.054908} {\bibfield  {journal} {\bibinfo
  {journal} {Phys. Rev.}\ }\textbf {\bibinfo {volume} {C81}},\ \bibinfo {pages}
  {054908} (\bibinfo {year} {2010})},\ \Eprint
  {https://arxiv.org/abs/0909.1717} {arXiv:0909.1717 [nucl-ex]} \BibitemShut
  {NoStop}%
%%CITATION = ARXIV:0909.1717;%%
\bibitem [{\citenamefont {Adamczyk}\ \emph
  {et~al.}(2014{\natexlab{a}})\citenamefont {Adamczyk} \emph
  {et~al.}}]{Adamczyk:2014mzf}%
  \BibitemOpen
  \bibfield  {author} {\bibinfo {author} {\bibfnamefont {L.}~\bibnamefont
  {Adamczyk}} \emph {et~al.} (\bibinfo {collaboration} {STAR}),\ }\href
  {https://doi.org/10.1103/PhysRevLett.113.052302} {\bibfield  {journal}
  {\bibinfo  {journal} {Phys. Rev. Lett.}\ }\textbf {\bibinfo {volume} {113}},\
  \bibinfo {pages} {052302} (\bibinfo {year} {2014}{\natexlab{a}})},\ \Eprint
  {https://arxiv.org/abs/1404.1433} {arXiv:1404.1433 [nucl-ex]} \BibitemShut
  {NoStop}%
%%CITATION = ARXIV:1404.1433;%%
\bibitem [{\citenamefont {Sirunyan}\ \emph {et~al.}(2018)\citenamefont
  {Sirunyan} \emph {et~al.}}]{Sirunyan:2017quh}%
  \BibitemOpen
  \bibfield  {author} {\bibinfo {author} {\bibfnamefont {A.~M.}\ \bibnamefont
  {Sirunyan}} \emph {et~al.} (\bibinfo {collaboration} {CMS}),\ }\href
  {https://doi.org/10.1103/PhysRevC.97.044912} {\bibfield  {journal} {\bibinfo
  {journal} {Phys. Rev.}\ }\textbf {\bibinfo {volume} {C97}},\ \bibinfo {pages}
  {044912} (\bibinfo {year} {2018})},\ \Eprint
  {https://arxiv.org/abs/1708.01602} {arXiv:1708.01602 [nucl-ex]} \BibitemShut
  {NoStop}%
%%CITATION = ARXIV:1708.01602;%%
\bibitem [{\citenamefont {Acharya}\ \emph {et~al.}(2018)\citenamefont {Acharya}
  \emph {et~al.}}]{Acharya:2017fau}%
  \BibitemOpen
  \bibfield  {author} {\bibinfo {author} {\bibfnamefont {S.}~\bibnamefont
  {Acharya}} \emph {et~al.} (\bibinfo {collaboration} {ALICE}),\ }\href
  {https://doi.org/10.1016/j.physletb.2017.12.021} {\bibfield  {journal}
  {\bibinfo  {journal} {Phys. Lett.}\ }\textbf {\bibinfo {volume} {B777}},\
  \bibinfo {pages} {151} (\bibinfo {year} {2018})},\ \Eprint
  {https://arxiv.org/abs/1709.04723} {arXiv:1709.04723 [nucl-ex]} \BibitemShut
  {NoStop}%
%%CITATION = ARXIV:1709.04723;%%
\bibitem [{\citenamefont {Xu}\ \emph {et~al.}(2018)\citenamefont {Xu},
  \citenamefont {Wang}, \citenamefont {Li}, \citenamefont {Zhao}, \citenamefont
  {Lin}, \citenamefont {Shen},\ and\ \citenamefont {Wang}}]{Xu:2017zcn}%
  \BibitemOpen
  \bibfield  {author} {\bibinfo {author} {\bibfnamefont {H.-J.}\ \bibnamefont
  {Xu}}, \bibinfo {author} {\bibfnamefont {X.}~\bibnamefont {Wang}}, \bibinfo
  {author} {\bibfnamefont {H.}~\bibnamefont {Li}}, \bibinfo {author}
  {\bibfnamefont {J.}~\bibnamefont {Zhao}}, \bibinfo {author} {\bibfnamefont
  {Z.-W.}\ \bibnamefont {Lin}}, \bibinfo {author} {\bibfnamefont
  {C.}~\bibnamefont {Shen}},\ and\ \bibinfo {author} {\bibfnamefont
  {F.}~\bibnamefont {Wang}},\ }\href
  {https://doi.org/10.1103/PhysRevLett.121.022301} {\bibfield  {journal}
  {\bibinfo  {journal} {Phys. Rev. Lett.}\ }\textbf {\bibinfo {volume} {121}},\
  \bibinfo {pages} {022301} (\bibinfo {year} {2018})},\ \Eprint
  {https://arxiv.org/abs/1710.03086} {arXiv:1710.03086 [nucl-th]} \BibitemShut
  {NoStop}%
%%CITATION = ARXIV:1710.03086;%%
\bibitem [{\citenamefont {Adamczyk}\ \emph
  {et~al.}(2014{\natexlab{b}})\citenamefont {Adamczyk} \emph
  {et~al.}}]{Adamczyk:2013kcb}%
  \BibitemOpen
  \bibfield  {author} {\bibinfo {author} {\bibfnamefont {L.}~\bibnamefont
  {Adamczyk}} \emph {et~al.} (\bibinfo {collaboration} {STAR}),\ }\href
  {https://doi.org/10.1103/PhysRevC.89.044908} {\bibfield  {journal} {\bibinfo
  {journal} {Phys. Rev.}\ }\textbf {\bibinfo {volume} {C89}},\ \bibinfo {pages}
  {044908} (\bibinfo {year} {2014}{\natexlab{b}})},\ \Eprint
  {https://arxiv.org/abs/1303.0901} {arXiv:1303.0901 [nucl-ex]} \BibitemShut
  {NoStop}%
%%CITATION = ARXIV:1303.0901;%%
\bibitem [{\citenamefont {Wang}\ and\ \citenamefont
  {Zhao}(2017)}]{Wang:2016iov}%
  \BibitemOpen
  \bibfield  {author} {\bibinfo {author} {\bibfnamefont {F.}~\bibnamefont
  {Wang}}\ and\ \bibinfo {author} {\bibfnamefont {J.}~\bibnamefont {Zhao}},\
  }\href {https://doi.org/10.1103/PhysRevC.95.051901} {\bibfield  {journal}
  {\bibinfo  {journal} {Phys. Rev.}\ }\textbf {\bibinfo {volume} {C95}},\
  \bibinfo {pages} {051901} (\bibinfo {year} {2017})},\ \Eprint
  {https://arxiv.org/abs/1608.06610} {arXiv:1608.06610 [nucl-th]} \BibitemShut
  {NoStop}%
%%CITATION = ARXIV:1608.06610;%%
\bibitem [{\citenamefont {Ajitanand}\ \emph {et~al.}(2011)\citenamefont
  {Ajitanand}, \citenamefont {Lacey}, \citenamefont {Taranenko},\ and\
  \citenamefont {Alexander}}]{Ajitanand:2010rc}%
  \BibitemOpen
  \bibfield  {author} {\bibinfo {author} {\bibfnamefont {N.~N.}\ \bibnamefont
  {Ajitanand}}, \bibinfo {author} {\bibfnamefont {R.~A.}\ \bibnamefont
  {Lacey}}, \bibinfo {author} {\bibfnamefont {A.}~\bibnamefont {Taranenko}},\
  and\ \bibinfo {author} {\bibfnamefont {J.~M.}\ \bibnamefont {Alexander}},\
  }\href {https://doi.org/10.1103/PhysRevC.83.011901} {\bibfield  {journal}
  {\bibinfo  {journal} {Phys. Rev.}\ }\textbf {\bibinfo {volume} {C83}},\
  \bibinfo {pages} {011901} (\bibinfo {year} {2011})},\ \Eprint
  {https://arxiv.org/abs/1009.5624} {arXiv:1009.5624 [nucl-ex]} \BibitemShut
  {NoStop}%
%%CITATION = ARXIV:1009.5624;%%
\bibitem [{\citenamefont {Bzdak}(2012)}]{Bzdak:2011np}%
  \BibitemOpen
  \bibfield  {author} {\bibinfo {author} {\bibfnamefont {A.}~\bibnamefont
  {Bzdak}},\ }\href {https://doi.org/10.1103/PhysRevC.85.044919} {\bibfield
  {journal} {\bibinfo  {journal} {Phys. Rev.}\ }\textbf {\bibinfo {volume}
  {C85}},\ \bibinfo {pages} {044919} (\bibinfo {year} {2012})},\ \Eprint
  {https://arxiv.org/abs/1112.4066} {arXiv:1112.4066 [nucl-th]} \BibitemShut
  {NoStop}%
%%CITATION = ARXIV:1112.4066;%%
\bibitem [{\citenamefont {Zhao}\ \emph {et~al.}(2017)\citenamefont {Zhao},
  \citenamefont {Li},\ and\ \citenamefont {Wang}}]{Zhao:2017nfq}%
  \BibitemOpen
  \bibfield  {author} {\bibinfo {author} {\bibfnamefont {J.}~\bibnamefont
  {Zhao}}, \bibinfo {author} {\bibfnamefont {H.}~\bibnamefont {Li}},\ and\
  \bibinfo {author} {\bibfnamefont {F.}~\bibnamefont {Wang}},\ }\href@noop {}
  {\  (\bibinfo {year} {2017})},\ \Eprint {https://arxiv.org/abs/1705.05410}
  {arXiv:1705.05410 [nucl-ex]} \BibitemShut {NoStop}%
%%CITATION = ARXIV:1705.05410;%%
\bibitem [{\citenamefont {Bloczynski}\ \emph {et~al.}(2013)\citenamefont
  {Bloczynski}, \citenamefont {Huang}, \citenamefont {Zhang},\ and\
  \citenamefont {Liao}}]{Bloczynski:2012en}%
  \BibitemOpen
  \bibfield  {author} {\bibinfo {author} {\bibfnamefont {J.}~\bibnamefont
  {Bloczynski}}, \bibinfo {author} {\bibfnamefont {X.-G.}\ \bibnamefont
  {Huang}}, \bibinfo {author} {\bibfnamefont {X.}~\bibnamefont {Zhang}},\ and\
  \bibinfo {author} {\bibfnamefont {J.}~\bibnamefont {Liao}},\ }\href
  {https://doi.org/10.1016/j.physletb.2012.12.030} {\bibfield  {journal}
  {\bibinfo  {journal} {Phys. Lett.}\ }\textbf {\bibinfo {volume} {B718}},\
  \bibinfo {pages} {1529} (\bibinfo {year} {2013})},\ \Eprint
  {https://arxiv.org/abs/1209.6594} {arXiv:1209.6594 [nucl-th]} \BibitemShut
  {NoStop}%
%%CITATION = ARXIV:1209.6594;%%
\bibitem [{\citenamefont {Bloczynski}\ \emph {et~al.}(2015)\citenamefont
  {Bloczynski}, \citenamefont {Huang}, \citenamefont {Zhang},\ and\
  \citenamefont {Liao}}]{Bloczynski:2013mca}%
  \BibitemOpen
  \bibfield  {author} {\bibinfo {author} {\bibfnamefont {J.}~\bibnamefont
  {Bloczynski}}, \bibinfo {author} {\bibfnamefont {X.-G.}\ \bibnamefont
  {Huang}}, \bibinfo {author} {\bibfnamefont {X.}~\bibnamefont {Zhang}},\ and\
  \bibinfo {author} {\bibfnamefont {J.}~\bibnamefont {Liao}},\ }\href
  {https://doi.org/10.1016/j.nuclphysa.2015.03.012} {\bibfield  {journal}
  {\bibinfo  {journal} {Nucl. Phys. A}\ }\textbf {\bibinfo {volume} {939}},\
  \bibinfo {pages} {85} (\bibinfo {year} {2015})},\ \Eprint
  {https://arxiv.org/abs/1311.5451} {arXiv:1311.5451 [nucl-th]} \BibitemShut
  {NoStop}%
\bibitem [{\citenamefont {Zhao}\ \emph {et~al.}(2018)\citenamefont {Zhao},
  \citenamefont {Ma},\ and\ \citenamefont {Ma}}]{Zhao_2018}%
  \BibitemOpen
  \bibfield  {author} {\bibinfo {author} {\bibfnamefont {X.-L.}\ \bibnamefont
  {Zhao}}, \bibinfo {author} {\bibfnamefont {Y.-G.}\ \bibnamefont {Ma}},\ and\
  \bibinfo {author} {\bibfnamefont {G.-L.}\ \bibnamefont {Ma}},\ }\bibfield
  {journal} {\bibinfo  {journal} {Physical Review C}\ }\textbf {\bibinfo
  {volume} {97}},\ \href {https://doi.org/10.1103/physrevc.97.024910}
  {10.1103/physrevc.97.024910} (\bibinfo {year} {2018})\BibitemShut {NoStop}%
\bibitem [{\citenamefont {Belmont}\ and\ \citenamefont
  {Nagle}(2017)}]{Belmont:2016oqp}%
  \BibitemOpen
  \bibfield  {author} {\bibinfo {author} {\bibfnamefont {R.}~\bibnamefont
  {Belmont}}\ and\ \bibinfo {author} {\bibfnamefont {J.~L.}\ \bibnamefont
  {Nagle}},\ }\href {https://doi.org/10.1103/PhysRevC.96.024901} {\bibfield
  {journal} {\bibinfo  {journal} {Phys. Rev. C}\ }\textbf {\bibinfo {volume}
  {96}},\ \bibinfo {pages} {024901} (\bibinfo {year} {2017})},\ \Eprint
  {https://arxiv.org/abs/1610.07964} {arXiv:1610.07964 [nucl-th]} \BibitemShut
  {NoStop}%
\bibitem [{\citenamefont {Khachatryan}\ \emph {et~al.}(2017)\citenamefont
  {Khachatryan} \emph {et~al.}}]{Khachatryan:2016got}%
  \BibitemOpen
  \bibfield  {author} {\bibinfo {author} {\bibfnamefont {V.}~\bibnamefont
  {Khachatryan}} \emph {et~al.} (\bibinfo {collaboration} {CMS}),\ }\href
  {https://doi.org/10.1103/PhysRevLett.118.122301} {\bibfield  {journal}
  {\bibinfo  {journal} {Phys. Rev. Lett.}\ }\textbf {\bibinfo {volume} {118}},\
  \bibinfo {pages} {122301} (\bibinfo {year} {2017})},\ \Eprint
  {https://arxiv.org/abs/1610.00263} {arXiv:1610.00263 [nucl-ex]} \BibitemShut
  {NoStop}%
%%CITATION = ARXIV:1610.00263;%%
\bibitem [{\citenamefont {Deng}\ \emph {et~al.}(2016)\citenamefont {Deng},
  \citenamefont {Huang}, \citenamefont {Ma},\ and\ \citenamefont
  {Wang}}]{Deng:2016knn}%
  \BibitemOpen
  \bibfield  {author} {\bibinfo {author} {\bibfnamefont {W.-T.}\ \bibnamefont
  {Deng}}, \bibinfo {author} {\bibfnamefont {X.-G.}\ \bibnamefont {Huang}},
  \bibinfo {author} {\bibfnamefont {G.-L.}\ \bibnamefont {Ma}},\ and\ \bibinfo
  {author} {\bibfnamefont {G.}~\bibnamefont {Wang}},\ }\href
  {https://doi.org/10.1103/PhysRevC.94.041901} {\bibfield  {journal} {\bibinfo
  {journal} {Phys. Rev.}\ }\textbf {\bibinfo {volume} {C94}},\ \bibinfo {pages}
  {041901} (\bibinfo {year} {2016})},\ \Eprint
  {https://arxiv.org/abs/1607.04697} {arXiv:1607.04697 [nucl-th]} \BibitemShut
  {NoStop}%
%%CITATION = ARXIV:1607.04697;%%
\bibitem [{\citenamefont {Miller}(2010)}]{Miller:2010nz}%
  \BibitemOpen
  \bibfield  {author} {\bibinfo {author} {\bibfnamefont {G.~A.}\ \bibnamefont
  {Miller}},\ }\href {https://doi.org/10.1146/annurev.nucl.012809.104508}
  {\bibfield  {journal} {\bibinfo  {journal} {Ann. Rev. Nucl. Part. Sci.}\
  }\textbf {\bibinfo {volume} {60}},\ \bibinfo {pages} {1} (\bibinfo {year}
  {2010})},\ \Eprint {https://arxiv.org/abs/1002.0355} {arXiv:1002.0355
  [nucl-th]} \BibitemShut {NoStop}%
%%CITATION = ARXIV:1002.0355;%%
\bibitem [{\citenamefont {Alberico}\ \emph {et~al.}(2009)\citenamefont
  {Alberico}, \citenamefont {Bilenky}, \citenamefont {Giunti},\ and\
  \citenamefont {Graczyk}}]{Alberico:2008sz}%
  \BibitemOpen
  \bibfield  {author} {\bibinfo {author} {\bibfnamefont {W.~M.}\ \bibnamefont
  {Alberico}}, \bibinfo {author} {\bibfnamefont {S.~M.}\ \bibnamefont
  {Bilenky}}, \bibinfo {author} {\bibfnamefont {C.}~\bibnamefont {Giunti}},\
  and\ \bibinfo {author} {\bibfnamefont {K.~M.}\ \bibnamefont {Graczyk}},\
  }\href {https://doi.org/10.1103/PhysRevC.79.065204} {\bibfield  {journal}
  {\bibinfo  {journal} {Phys. Rev.}\ }\textbf {\bibinfo {volume} {C79}},\
  \bibinfo {pages} {065204} (\bibinfo {year} {2009})},\ \Eprint
  {https://arxiv.org/abs/0812.3539} {arXiv:0812.3539 [hep-ph]} \BibitemShut
  {NoStop}%
%%CITATION = ARXIV:0812.3539;%%
\bibitem [{\citenamefont {Deng}\ \emph {et~al.}(2012)\citenamefont {Deng},
  \citenamefont {Xu},\ and\ \citenamefont {Greiner}}]{Deng:2011at}%
  \BibitemOpen
  \bibfield  {author} {\bibinfo {author} {\bibfnamefont {W.-T.}\ \bibnamefont
  {Deng}}, \bibinfo {author} {\bibfnamefont {Z.}~\bibnamefont {Xu}},\ and\
  \bibinfo {author} {\bibfnamefont {C.}~\bibnamefont {Greiner}},\ }\href
  {https://doi.org/10.1016/j.physletb.2012.04.010} {\bibfield  {journal}
  {\bibinfo  {journal} {Phys. Lett.}\ }\textbf {\bibinfo {volume} {B711}},\
  \bibinfo {pages} {301} (\bibinfo {year} {2012})},\ \Eprint
  {https://arxiv.org/abs/1112.0470} {arXiv:1112.0470 [hep-ph]} \BibitemShut
  {NoStop}%
%%CITATION = ARXIV:1112.0470;%%
\end{thebibliography}%

\end{document}